\documentclass{elsarticle}

\usepackage{amssymb,amsmath}
\usepackage{url}
\usepackage{balance}

\usepackage{algorithmic}
\usepackage[lined,ruled,linesnumbered]{algorithm2e}

\newtheorem{proposition}{Proposition}
\DeclareMathOperator{\tr}{tr}
\DeclareMathOperator{\vect}{vec}
\DeclareMathOperator{\var}{var}

\usepackage{color}
\usepackage{tikz}
\usetikzlibrary{arrows.meta}
\definecolor{mycolor}{RGB}{189,215,238}

\newcommand{\bvartheta}{\boldsymbol{\vartheta}}
\newcommand{\by}{\boldsymbol{y}}
\newcommand{\bm}{\boldsymbol{m}}
\newcommand{\bC}{\boldsymbol{C}}
\newcommand{\bY}{\boldsymbol{Y}}

\newcommand{\esp}{\mathsf{E}} 
\newcommand{\bkappa}{\boldsymbol{\kappa}}
\newcommand{\btheta}{\boldsymbol{\theta}}
\newcommand{\bomega}{\boldsymbol{\omega}}
\newcommand{\bH}{\boldsymbol{H}}
\newcommand{\bh}{\boldsymbol{h}}
\newcommand{\bsH}{\boldsymbol{\mathsf{H}}}
\newcommand{\bsh}{\boldsymbol{\mathsf{h}}}
\newcommand{\bmu}{\boldsymbol{\mu}} 
\newcommand{\bR}{\boldsymbol{R}}
\newcommand{\bI}{\boldsymbol{I}}
\newcommand{\be}{\boldsymbol{e}}
\newcommand{\bzero}{\boldsymbol{0}}   
\newcommand{\bW}{\boldsymbol{W}}
\newcommand{\bb}{\boldsymbol{b}} 
\newcommand{\bT}{\boldsymbol{T}}
\newcommand{\bQ}{\boldsymbol{Q}}
\newcommand{\bZ}{\boldsymbol{Z}}
\newcommand{\bN}{\boldsymbol{N}}
\newcommand{\bc}{\boldsymbol{c}}
\newcommand{\bGamma}{\boldsymbol{\Gamma}}
\newcommand{\bDelta}{\boldsymbol{\Delta}}
\newcommand{\bS}{\boldsymbol{S}}

\newcommand{\bX}{\boldsymbol{X}}

\begin{document}

\title{Online change-point detection with kernels}

\author[1]{Andr{\'e} Ferrari\corref{cor1}}
\ead{Andre.Ferrari@univ-cotedazur.fr}

\author[1]{C{\'e}dric Richard}
\ead{Cedric.Richard@univ-cotedazur.fr}

\author[2]{Anthony Bourrier}
\ead{Anthony.Bourrier@thalesaleniaspace.com}

\author[1]{Ikram Bouchikhi}
\ead{Ikram.Bouchikhi@univ-cotedazur.fr}

\cortext[cor1]{Corresponding author}

\address[1]{Universit{\'e} C{\^o}te d'Azur, Observatoire de la C\^ote d'Azur, CNRS,  Lab. Lagrange, France}
\address[2]{Thales Alenia Space, Cannes la Bocca, France}

\begin{abstract}
Change-points in time series data are usually defined as the time instants at which changes in their properties occur. Detecting change-points is critical in a number of applications as diverse as detecting credit card and insurance frauds, or intrusions into networks. Recently the authors introduced an online kernel-based change-point detection method built upon direct estimation of the density ratio on consecutive time intervals. This paper further investigates this algorithm, making improvements and analyzing its behavior in the mean and mean square sense, in the absence and presence of a change point. These theoretical analyses are validated with Monte Carlo simulations. The detection performance of the algorithm is illustrated through experiments on real-world data and compared to state of the art methodologies.
\end{abstract}


\maketitle

\section{Introduction} 

From a statistical perspective, a change-point is defined as a time instant at which some properties of a signal change, that is, the observations belong to one state up to that point, and belong to an other state after it. This change can be caused by external events, as well as by sharp transitions in the dynamics of the signal, either way it can hold critical information.  Among possible applications of change point detection (CPD) we can mention medical monitoring~\cite{yuan2014automatic, gajic2015detection,d2015automated}, finance~\cite{bolton2002statistical} and network security~\cite{tartakovsky2006detection,yan2008catching}. We refer interested readers to, e.g., \cite{pimentel_review_2014} or \cite{truong_selective_2020} for comprehensive reviews of CPD algorithms.

CPD algorithms can be classified, based on what is assumed to be known about the data distribution, as \emph{parametric} or \emph{non-parametric}.  \emph{Parametric} approaches assume that a model describing the data distributions of the different states is available. For instance, cumulative sum (CUSUM) type algorithms~\cite{Bass93} assume, in their simplest form, that the parameter that undergoes changes is known, but also require knowledge of its pre-change and sometimes post-change values, e.g., change in the mean or in the variance~\cite{Inc94}. In case where the aforementioned parameters are unknown, the generalized likelihood ratio~\cite{Gust96}, which consists of substituting all the unknown parameters by their maximum likelihood estimates, can be used. Less restrictive approaches have also been devised. Among these, subspace identification techniques are built upon the idea that if, at a certain time instant, there is a change in the mechanism generating the time series, then the (linear) subspace spanned by the signal trajectory also changes. This principle is used in \cite{Kawahara08} where the authors explicitly model the observations via a discrete-time linear state-space system. Another example is the Singular Spectrum Transformation, which calculates distance-based change-point scores by comparing singular spectra of two trajectory matrices over consecutive windows~\cite{itoh2010change, moskvina2003algorithm}. If all assumptions about the data model are met, these techniques can be robust and efficient. In practice though, stochastic models that properly describe the data are not often available. And, even when they are, data are susceptible to deviations from the assumed models. \emph{Non-parametric} approaches were introduced to cope with these limitations. They can be used in a broader range of applications, since they do not require (strong) prior information. 

Non-parametric algorithms are usually classified as \emph{supervised} or \emph{unsupervised} methods. In the first case, when the number of possible states is specified, and labeled data representing each state is available, machine learning algorithms can be used to train multi-class classifiers and then find each state boundary. If not, the nominal-state sequences represent the unique class and the problem can be solved using, e.g., a one-class algorithm such as~\cite{Schlkopf2001} and~\cite{Ma2003} or alternative approaches such as the Hilbert-Schmidt Independence Criterion in~\cite{yamada2013change}. However, in many practical situations, labeled data is not available and \emph{unsupervised} algorithms that can adapt to different situations are required. This problem can be tackled by extending subspace identification techniques to non-linear subspaces using, e.g., nearest neighbors algorithms or, more generally, manifold learning methods, \cite{Chen2019,Xie2013}. An alternative approach consists of operating in a Reproducing Kernel Hilbert Space (RKHS) in order to extend the use of linear models and algorithms to nonlinear problems, \cite{Shawe04}. This strategy is used in \cite{Zaremba2013} which proposes an online implementation of the Maximum Mean Discrepancy (MMD) two-sample test based on the B-statistics \cite{Zaremba2013}. Note that \cite{Harchaoui09}, where change-points are detected using a Kernel Fisher Discriminant, and \cite{Keriven2020}, where the authors propose to monitor the mean of the process in the feature space, are both strongly related to MMD. Another class of \emph{unsupervised} methods is based on the direct estimation of the ratio of probability densities of the data over consecutive segments. They include the Kullback-Leibler Importance Estimation Procedure (KLIEP), the Unconstrained Least Squares Importance Fitting (uLSIF) and the Relative Unconstrained Least Squares Importance Fitting (RuLSIF)~\cite{Song13}. The main contributions of this article lie in this class of methods. 

Recently, an online version of a RuLSIF-based CPD algorithm, which consists of estimating the density ratio over consecutive intervals of the time series data, was introduced~\cite{Ikram2018}. In this algorithm, %
the model parameters are estimated in an online and adaptive way similar to the Kernel Least Mean Squares (KLMS) algorithm~\cite{Liu08}. The methodology showed promising and reliable detection results. In~\cite{Ikram2019} the authors proposed to modify the original cost function used in~\cite{Ikram2018} in order to further improve the performance and achieve unbiasedness of the algorithm, referred to as NOUGAT (Nonparametric Online chanGepoint detection AlgoriThm).

The main contribution of this paper are as follows. After introducing the proposed algorithm (Section \ref{The proposed algorithm}), we provide a theoretical analysis of its stochastic behavior by deriving models for the mean and the variance of the detection statistics, in the absence and the presence of a change-point (Section \ref{Theoretical analysis}). These models are useful for several purposes: i) to assess the detection performance of the algorithm; ii) for detector design and optimization. %
Then we demonstrate the accuracy of these models, and we present performance comparisons with state-of-the-art algorithms on simulated and real data sets (Section \ref{Simulations}). Finally we conclude this paper with recommendations for future research (Section~\ref{Conclusion}).

\section{The NOUGAT algorithm} 
\label{The proposed algorithm}

In this section, we formulate the CPD problem. We review the proposed method and the online algorithm denoted as NOUGAT. Then we introduce the detection statistic. Finally we briefly discuss related works.

\subsection{Problem formulation}

We aim at detecting change-points in the distribution of independent random variables $\{\by_t\}_{t\in \mathbb{N}}$,  $\by_t\in \mathbb{R}^k$, by estimating a model $g(\cdot)$ for $r(\by) - 1 $, where $r(\by) = p_\text{test}(\by)/p_\text{ref}(\by)$ is the density ratio between the probability density $p_\text{test}(\by)$ of the data on a test interval:
\begin{equation}
\label{eq:Ttest}
 \bY^{\text{test}}_t = ( \by_{t - (N_\text{test}-1)}, \ldots, \by_{t-1}, \by_{t}) \in \mathbb{R}^{k \times N_\text{test}} 
\end{equation}
and the probability density $p_\text{ref}(\by)$ of the data on a reference interval:
\begin{equation}
\label{eq:Ref}
 \bY^{\text{ref}}_t = ( \by_{t - (N_\text{ref} + N_\text{test}-1)}, ,  \ldots,  \by_{t - N_\text{test}}) \in \mathbb{R}^{k \times N_\text{ref}}
 \end{equation}
where $N_\text{test}$ and $N_\text{ref}$ are the number of samples in the test and reference intervals, respectively. Note that, contrary to RuLSIF \cite{Song13},  $r(\by) - 1$ is  preferred  to $r(\by)$ because it leads to an unbiased estimator under the no change-point hypothesis as we shall see later.

In the general case of a scalar time series $\{y_t\}_{t\in \mathbb{N}}$,  as commonly reported in the literature, we propose to proceed by considering
\begin{equation} \label{sigtovec}
	\by_{t} = (y_{t}, y_{t+1}, \ldots, y_{t+k-1})^\top\in  \mathbb{R}^{k}
\end{equation}
to take into account any dependence that may exist between successive $y_t$.

\subsection{Density-ratio estimation}
 
The problem addressed in this paper consists of estimating a model $g(\cdot)$ for $r(\by) - 1 $. It can be solved by fitting $g(\by)$ to $r(\by)-1$ with respect to the squared loss:
\begin{equation}
	\label{cost_init}
	\mathcal{C}(g) = \frac{1}{2}\esp_{p_\text{ref}(\by)}\{(r(\by) - 1 - g(\by))^{2}\}
\end{equation}
Note that, as in~\cite{Song13},  the expectation operator is defined with respect to the reference interval. By expanding~\eqref{cost_init} and then using $r(\by) p_\text{ref}(\by) = p_\text{test}(\by)$, we obtain:
\begin{equation}
	\label{cost}
	\mathcal{C}(g)=  \frac{1}{2}\esp_{p_\text{ref}(\by)}\{g^2(\by)\}-\esp_{p_\text{test}(\by)}\{g(\by)\} +\esp_{p_\text{test}(\by)}\{g(\by)\} + C
\end{equation}
where $C$ denotes a constant value.
Approximating the expected values in~\eqref{cost} by their empirical averages over the reference and test intervals data $\bY^{\text{ref}}_t$ and $\bY^{\text{test}}_t$ for any fixed $t$, leads to the following empirical optimization problem:
\begin{align}
	\min_{g\in\mathcal{H}}&\left(\frac{1}{2N_\text{ref}}\sum_{i= t-(N_\text{ref}+N_\text{test}-1)}^{t-N_\text{test}}\!\!g^2(\by_i)- \frac{1}{N_\text{test}}\sum_{i=t-(N_\text{test}-1)}^{t}\!\!g(\by_i)\right. \nonumber\\
	&+ \left. \frac{1}{N_\text{ref}}\sum_{i= t-(N_\text{ref}+N_\text{test}-1)}^{t-N_\text{test}}\!\!g(\by_i)+ \nu\, \Omega(\|g\|_{\mathcal{H}})	\right) \label{empirical-cost}
\end{align}
where $\mathcal{H}$ denotes an arbitrary reproducing kernel Hilbert space of real-valued functions on $\mathbb{R}$. Let $\kappa(\cdot\,,\cdot)$ be the reproducing kernel of~$\mathcal{H}$. The term $\nu\,\Omega(\|g\|_{\mathcal{H}})$ with $\nu \geq 0$ is a regularization term  added to promote smoothness of the solution. By virtue of the Representer Theorem \cite{Sch2000}, any function $g(\cdot)$ of~$\mathcal{H}$ that minimizes~\eqref{empirical-cost} can be expressed as a kernel expansion in terms of available data:
\begin{equation}
 	g(\cdot, \btheta) = \sum_{i=t-(N_\text{ref} + N_\text{test} - 1)}^{t} {\theta}_i ~ \kappa(\cdot\,,\by_i) \label{mod_g}%
\end{equation}	
where the $\theta _{i}$ are parameters to be learned. This model cannot be trained efficiently in an online framework, as it needs to update both $\{\by_i\}$ and $\btheta$ as time $t$ progresses. A standard strategy in the literature is to substitute $\{\by_i\}$ in \eqref{mod_g} by a fixed dictionary of size $L$, $\{\by_{\omega_i}\}^L_{i=1}$, whose elements are chosen according to some sparsification rule~\cite{rojo17dspkm} to represent the input data space accurately, resulting in a fixed order model,
\begin{equation}
	\label{mod_r}
	g(\cdot, \btheta) = \sum_{i=1}^{L} {\theta}_i \kappa_{\omega_i}(\cdot)
\end{equation}
where $\kappa_{\omega_i}(\cdot) = \kappa(\cdot, \by_{\omega_i})$, for all $i\in\{1,\ldots,L\}$, are the elements of the  dictionary, and $\bkappa_{\bomega}(\cdot)= [\kappa_{\omega_1}(\cdot), \ldots, \kappa_{\omega_L}(\cdot)]^\top$. 

Substituting~\eqref{mod_r} into~\eqref{empirical-cost}, assuming a ridge parameter space regularization~\cite{Sch2000}, and minimizing~\eqref{empirical-cost} w.r.t. $\btheta$, we find that the optimal parameter vector $\hat{\btheta}_t$ is the solution of the following strictly convex quadratic optimization problem:
\begin{equation}
	\label{opt}
\begin{aligned}
		&\hat{\btheta}_t =\arg\min_{\btheta \in \mathbb{R}^{L}} J_{t}(\btheta)\\
		&\text{with }\,\, J_t(\btheta)=\frac{1}{2} \btheta^\top\bH_t^\text{ref} \btheta + \btheta^{\top} \be^\circ_t + \frac{\nu}{2} \|\btheta\|^{2} 
\end{aligned}
\end{equation}
where 
\begin{equation}
\label{e_opt}
\be^\circ_t = \bh_{t}^\text{ref} - \bh_{t}^\text{test}
\end{equation}
and
\begin{align}
    &\bh_{t}^\text{test}= \frac{1}{N_\text{test}} \sum_{i={t - (N_\text{test}-1)}}^{t} \bkappa_{\bomega}(\by_{i}) \label{h_test}\\
    &\bh_{t}^\text{ref}= \frac{1}{N_\text{ref}} \sum_{i={t - (N_\text{test} + N_\text{ref}-1)}}^{t - N_\text{test}} \bkappa_{\bomega}(\by_{i}) \label{h_ref}\\
 &\bH_{t}^\text{ref}= \frac{1}{N_\text{ref}} \sum_{i={t - (N_\text{test} + N_\text{ref}-1)}}^{t - N_\text{test}} \bkappa_{\bomega}(\by_{i})  \bkappa_{\bomega}^\top(\by_{i}) \label{H_ref}
\end{align}

\subsection{Online density-ratio estimation}
\label{sec:online}
Let $\btheta_{t} $ be an estimate of the parameter vector of the density ratio model at time instant $t$. When ${t}\rightarrow {t}+1$, according to~\eqref{opt}, $\btheta_{t+1}$ should be computed, as proposed in RuLSIF~\cite{Song13}, by updating first \eqref{h_test}--\eqref{H_ref} and then minimizing the updated criterion $J_{t+1}(\btheta)$. In order to reduce the computational cost, we propose as an alternative strategy to compute $\btheta_{t+1}$ by updating $\btheta_{t}$ based on a gradient descent step of $J_{t+1}(\btheta)$:
 \begin{align}
	\btheta_{t+1} &= \btheta_{t} - \mu \nabla J_{t+1}(\btheta_{{t}})	\label{grad_step} \\
	& =\btheta_{t}- \mu \big[(\bH_{t+1}^\text{ref} + \nu \bI)\btheta_{t}+ \be_{t+1}^\circ\big] \label{update_theta}
 \end{align}
where $\mu>0$ is a small step size, and $\nabla J_{t+1}(\btheta_t)$ denotes the gradient of $J_{t+1}(\cdot)$ evaluated at $\btheta_t$. The resulting algorithm shares similarities with the KLMS algorithm~\cite{Richard09}. The convergence behavior of the KLMS was analyzed in the case of a fixed dictionary in~\cite{Jie14}, and in a more general case in~\cite{Parreira12}. Additional constraints such as sparsity have been also considered~\cite{gao2013kernel}.

In practice, updating model $g(\cdot,\btheta_{t})$ at each time instant $t$ is a two-stage process that consists of updating both the dictionary $\{\by_{\omega_i}\}^L_{i=1}$ and the order $L$ of the kernel expansion~\eqref{mod_r}, followed by the update of parameter vector $\btheta_{t}$.

 \subsection{Dictionary update}
 
 Numerous strategies of dictionary learning have been introduced in the online kernel filtering literature. They consist of building the dictionary $\{\by_{\omega_i}\}^L_{i=1}$ sequentially, by inserting selected samples $\by_{i}$ that improve the representation of input data according to some criterion.  For instance, the Approximate Linear Dependency (ALD)~\cite{Engel04} criterion checks whether, in feature space $\mathcal{H}$, the new candidate element $\kappa(\cdot\,,\by_{t+1})$ can be well approximated by a linear combination of the elements $\kappa(\cdot\,,\by_{\omega_i})$ which are already in the dictionary. If not, it is added to the dictionary. The coherence rule \cite{Richard09} was introduced to avoid the computational complexity inherent to ALD. It is now considered as a state-of-the-art strategy and widely used as such. Defined by:
\begin{equation*}
	\eta = \max_{i\neq j} |\kappa(\by_{\omega_i}, \by_{\omega_j})|,
\end{equation*}
coherence $\eta$ reflects the largest correlation between the dictionary elements. The coherence rule for kernel-based dictionary selection consists of inserting $\by_{t+1}$ in dictionary $\{\by_{\bomega_i}\}^L_{i=1}$ provided that its coherence remains below a threshold $\eta_0$ preset by the user:
\begin{equation}\label{update dictionary}
    \max_{\by_{\omega_i} \in \{\by_{\bomega_i}\}^L_{i=1}} \ | \kappa(\by_{t+1}, \by_{\omega_i})| \leq \eta_0
\end{equation}
In \cite{Richard09} the authors show that the dimension of dictionaries determined with rule \eqref{update dictionary} is finite due to the compactness of the input space.

\subsection{NOUGAT Algorithm}

\label{Weights update}

Depending on whether the new sample $\by_{t+1}$ has been inserted into the dictionary, or not, parameter vector $\btheta_t$ is updated similarly to~\cite{Richard09}. At each time instant $t$, given $\btheta_{t}$, we propose as a test statistic to consider the average of the (shifted by 1) density ratio estimators over the test interval, namely: 
\begin{equation}
  g_t = \frac{1}{N_\text{test}} \sum_{i={t - (N_\text{test}-1)}}^{t} g(\by_{i}, \btheta_t) =\btheta_{t}^\top\bh^{\text{test}}_t\label{test_stat}
\end{equation}
CPD is then performed by comparing $g_t+1$ to a given threshold $\xi$. The corresponding NOUGAT algorithm is described in Alg. \ref{nougat_algo}.

\begin{algorithm}
	\begin{algorithmic}[1]
	\STATE Step size $\mu$, initial dictionary $\bomega$, regularization $\nu$, thresholds $\eta_0$ and $\xi$
	\FOR{$t=1,2,\ldots$}
	\STATE update $\bH_{t}^\text{ref}$, $\bh_{t}^\text{test}$ and $\be_{t}^\circ$ using \eqref{e_opt}-\eqref{H_ref}
	\IF{$\max_{\by_{\omega_i} \in \{\by_{\omega_i}\}^L_{i=1}}|\kappa(\by_{t+1}, \by_{\omega_i})|> \eta_0$}
	\STATE {\it \# the dictionary remains unchanged and $\btheta_{t}$ is updated using \eqref{update_theta}}
	\STATE {\begin{equation*}
		\btheta_{t+1} = \btheta_{t}- \mu \big[(\bH_{t+1}^\text{ref} + \nu \bI)\btheta_{t}+ \be_{t+1}^\circ\big]
			\end{equation*}}
	\ELSE 
	\STATE {\it \# $\by_{t+1}$ is added to the dictionary and $\btheta_{t}$ is updated}
	\STATE $L\leftarrow L+1$, $\omega_{L+1}=t+1$
	\STATE{\begin{equation*}
		\btheta_{t+1} = \left(
		\begin{array}{c}
		\btheta_{t}\\0	
		\end{array}
		\right)
		- \mu \left[(\bH_{t+1}^\text{ref} + \nu \bI)\left(
		\begin{array}{c}
		\btheta_{t}\\0	
		\end{array}
		\right)+ \be^\circ_{t+1} \right]
	\end{equation*}}
	\ENDIF
	\STATE {\it \# compute the test statistic and test}
	\STATE {$g_{t+1} = \btheta_{t+1}^\top\bh^{\text{test}}_{t+1}$}
	\IF{$|g_{t+1}+1| > \xi$}
	\STATE{flag $t+1$ as a change point }
	\ENDIF
	\ENDFOR
	\end{algorithmic}
	\caption{NOUGAT Algorithm}
	\label{nougat_algo}
\end{algorithm}

\subsection{Related works}
\label{Related work}

Iteration \eqref{update_theta} turns out to be related to the classical Geometric Moving Average algorithm (GMA) proposed in \cite{Roberts1959}. GMA monitors a geometrically weighted estimate of the mean of $\by_t$ and detects a change when the estimated mean deviates from its nominal value. Without loss of generality, the mean in the observation space can be replaced by the mean $\esp\{\bkappa_{\bomega}(\by)\}$ in the feature space defined by mapping $\bkappa_{\bomega}(\cdot)$, leading to:
\begin{equation}\label{GMA}
	\bvartheta_{t+1} = (1-\alpha)\bvartheta_{t} + \alpha \bkappa_{\bomega}(\by_{t+1})
\end{equation}
However, as pointed out in~\cite{Keriven2020}, a drawback of GMA is that it requires to know the nominal value of $\esp\{\bkappa_{\bomega}(\by)\}$ in order to be able to calculate the associated test statistic: $\|\bvartheta_{t} - \esp\{\bkappa_{\bomega}(\by)\}\|_2$.  

To solve this problems in the GMA framework, a natural approach consists of comparing the estimates of $\esp\{\bkappa_{\bomega}(\by_{t})\}$ on two sliding windows, namely, the reference interval~\eqref{eq:Ref} and the test interval~\eqref{eq:Ttest}, as proposed in the Moving Average (MA) algorithm  described in \cite{Keriven2020} which tracks: 
\begin{equation}
\|\be^\circ_{t}\|_2 = \|\bh^\text{test}_{t} - \bh^\text{ref}_{t} \|_2 \label{MA}
\end{equation}
The approach implemented by NOUGAT differs in so far as, instead of calculating a deviation between two quantities estimated over the test and reference intervals, it estimates a unique statistic $r(\by)$ over the two intervals which is inherently equal to~1 under the null hypothesis. Note that an alternative approach, called NEWMA, proposed recently in~\cite{Keriven2020}, consists of testing the deviation between two GMA with different forgetting factors. The GMA with the smallest forgetting factor is used to provide an estimation of the in-control quantity. Contrarily to NOUGAT these algorithms do not explicitly take into account the covariance of the data in the feature space. From this point of view NOUGAT is similar to the Kernel Fisher Discriminant Ratio \cite{Harchaoui09} but with a much lower computation footprint since it does not require the inversion of a covariance matrix for each $t$. Concerning the memory resources, NOUGAT requires to buffer $N_\text{ref}+N_\text{test}$ data points. The B-statistics CPD algorithm \cite{Zaremba2013} also relies on a single sliding test window, but, conversely, on multiple reference windows which considerably increases the memory requirement.

In Section \ref{Performances comparison} we shall compare the detection performance and computational load of the three algorithms mentioned above with the same memory footprint, namely, MA, NOUGAT and RuLSIF.

\section{Theoretical analysis} \label{Theoretical analysis}

In this section we analyze the stochastic behavior of the proposed algorithm, and derive conditions for its stability in the mean and mean square sense, in the absence and presence of a change-point. To make the analysis tractable, we shall conduct it in the case of a \emph{pre-tuned} dictionary, i.e., a fixed dictionary of size $L$ is assumed to be available beforehand. This means that $L$ is fixed and the $\{\by_{\omega_i}\}^L_{i=1}$ are assumed to be deterministic.
The classical Modified Independence Assumption~(MIA) \cite{minkov2001}, which assumes that $\bH_{t+1}^\text{ref}$ and $\btheta_t$ are statistically independent, will also be considered. Although not true in general, this assumption is commonly used to analyze adaptive constructions since it allows to simplify the derivations without constraining the conclusions. There are several results in the adaptation literature that show that performance results that are obtained under this assumption match well the actual performance of the algorithms when the step-size is sufficiently small.

Using the update rule \eqref{update_theta}, we obtain the following recursion for $\btheta_{t}$:
\begin{equation}
\btheta_{t+1} =  \big[\bI - \mu(\bH_{t+1}^\text{ref} + \nu \bI) \big]\btheta_{t}   - 
\mu \be^\circ_{t+1} \label{weight_error} 
\end{equation}
Define:
\begin{align}
&\bsh^\text{test}_t = \esp_{p_\text{test}(\by)}\{\bh_{t}^\text{test}\} \label{esp_ht}\\
	&\bsh^\text{ref}_t = \esp_{p_\text{ref}(\by)}\{\bh_{t}^\text{ref}\} \label{esp_hr}\\
	&\bsH^\text{ref}_t = \esp_{p_\text{ref}(\by)}\{\bH_{t}^\text{ref}\} \label{esp_H}
\end{align}

Taking the expected values on both sides of~\eqref{weight_error} and using the MIA we obtain the mean weight model:
\begin{equation}
\bm_{\btheta, t+1}=  \big[\bI - \mu(\bsH^\text{ref}_{t+1}+ \nu \bI) \big]~ \bm_{\btheta, t} + \mu (\bsh^\text{test}_{t+1} - \bsh^\text{ref}_{t+1})\label{weight_error_bias} 
\end{equation}

We denote by $\bC_{\btheta,t}$ the correlation matrix of the weight vector $\btheta_t$:
\begin{equation*}
    \bC_{\btheta,t} =  \esp \{\btheta_t \btheta_t^\top \}
\end{equation*}
Estimating the variance of the test statistics requires a model for matrix $\bC_{\btheta,t}$. Post-multiplying \eqref{weight_error} by its transpose, taking the expectation, and using the MIA, we obtain the following recursive expression:
\begin{align}
	\bC_{\btheta,t+1}\!&=\!(1\!-\!\mu\nu)^2 \bC_{\theta,t} - \mu(1\!-\!\mu\nu) (\bsH^\text{ref}_{t+1}  \bC_{\btheta,t}  + \bC_{\btheta,t} \bsH^\text{ref}_{t+1}) \nonumber\\
	&+ \mu^2 (\bT\! +\! \bQ\! +\! \bZ\! +\! \bZ^\top\!) \!- \mu(1-\mu\nu) (\bN\!+\! \bN^\top) \label{iterC_H1}
\end{align}
where:
\begin{align}
    &\bT = \esp \{\bH_{t+1}^\text{ref} \btheta_t \btheta_t^\top \bH_{t+1}^\text{ref}\} \label{defT} \\
	&\bQ = \esp \{ \be^\circ_{t+1}{\be^\circ_{t+1}}^\top\} \label{defQ}\\
	&\bZ = \esp \{\be^\circ_{t+1} \btheta_t^\top \bH_{t+1}^\text{ref}\} \label{defZ} \\
    &\bN = \esp \{\be^\circ_{t+1} \btheta_t^\top\}  \label{defN} 
\end{align}
In the general, all these matrices can depend on $t$. To simplify the notations, this dependence is dropped.

\subsection{Stochastic behavior analysis under the null hypothesis}
\label{sec:Stochastic behavior analysis under the null}

\subsubsection{Mean analysis} \label{mean analysis}

Under the null hypothesis we have:
\begin{align*}
&\bsh^\text{ref}_t = \bsh^\text{test}_t = \esp_{p_\text{ref}(\by)}\{\bkappa_{\bomega}(\by)\}= \esp_{p_\text{test}(\by)}\{\bkappa_{\bomega}(\by)\}=\bh\\
&\bsH^\text{ref}_t=\esp_{p_\text{ref}(\by)}\{\bkappa_{\bomega}(\by)~\bkappa_{\bomega}^\top(\by)\}=\bH
\end{align*}
and the mean weight model~\eqref{weight_error_bias} simplifies to:
\begin{equation}
\bm_{\btheta, t+1} =  \big[\bI - \mu(\bH+ \nu \bI) \big]~\bm_{\btheta, t} \label{weight_mean_underNull} 
\end{equation}

The mean stability of the algorithm is then ensured by using a step size $\mu$ that satisfies:
\begin{equation}
	\mu< \frac{2}{\zeta_{\max}\{\bH+\nu\bI\}} \label{mu_max}
\end{equation}
where $\zeta_{\max}\{\cdot\}$ stands for the maximal eigenvalue of its matrix argument. Under this assumption $\bm_{\theta, t}\rightarrow \bzero$. When $\by$ is Gaussian distributed, analytical expressions of $\bh$ and $\bH$ for a Gaussian reproducing kernel can be derived; see  \ref{Computation of H and h}.

Taking the expectation of~\eqref{test_stat} and assuming that $\btheta_t$ and $\bh_{t}^\text{test}$ are independent, we get the mean of the test statistics~$g_t$:
\begin{equation}
    \esp \{g_t\} =  \bh^\top \bm_{\btheta, t}  \label{mean_detector}
\end{equation}
The necessary independence assumption together with the MIA will be validated by computer simulations.

Assuming \eqref{mu_max} holds, under the null hypothesis, the asymptotic unbiasedness of the estimator implies $\lim_{t\rightarrow\infty} \esp \{g_t\} = 0 $. When initializing \eqref{weight_error} with $\btheta_0=\bzero$, namely, $\bm_{\btheta, 0} =\bzero$, equation~\eqref{weight_mean_underNull} implies $\bm_{\btheta, t}=\bzero$ for all $t$. As a consequence $\esp \{g_t\}=0$, which means that the estimation of the density ratio $r(\by_t)=1 $ is unbiased under the null hypothesis for all~$t$. 

\subsubsection{Mean squared analysis}

The general model of $\bC_{\btheta,t}$ in \eqref{iterC_H1} depends on the matrices $\bT$, $\bQ$, $\bZ$ and $\bN$, defined in \eqref{defT}--\eqref{defN}. These matrices can be computed under the null hypothesis as follows. 
\begin{itemize}
\item Denoting $\bc_{\btheta,t} = \vect(\bC_{\btheta,t})$ where $\vect(\cdot)$ refers to the standard vectorization operator that stacks the columns of a matrix on top of each other, using the MIA and $\vect(\bf{ABC}) = (C^\top \otimes A)\vect(B)$ with $\otimes$ the Kronecker product, we find:
\begin{align}
\bT = \frac{1}{N_\text{ref}}~ \Big( \vect^{-1}(\bGamma \bc_{\btheta,t}) + (N_\text{ref}-1)~  \bH \bC_{\btheta,t} \bH\Big) \label{eq:T}
\end{align}
where $\bGamma$ is the $(L^2 \times L^2)$ matrix defined by:
\begin{equation} 
\bGamma = \esp_{p_\text{ref}(\by)} \{\bkappa_{\bomega}(\by)\bkappa_{\bomega}^\top(\by)  \otimes \bkappa_{\bomega}(\by)\bkappa_{\bomega}^\top(\by) \} \label{Gamma}
\end{equation}
The expression of $\bGamma$ is given in  \ref{Computation of Gamma and Delta}.

\item Under the null hypothesis, $\bQ$ is given by:
\begin{align}
	\bQ &= \frac{N_\text{ref}+ N_\text{test}}{N_\text{ref}~ N_\text{test}}~ ( \bH - \bh \bh^\top ) \label{eq:Q'}
\end{align}

\item In the same way as $\bT$, we find that:
\begin{align}
 \bZ &= \frac{1}{N_\text{ref}} \left(\vect^{-1} (\bDelta \bm_{\btheta,t}) - \bh \bm_{\btheta,t}^\top \bH \right) \label{eq:R'}
\end{align}
where $\bDelta$ is the $(L^2 \times L)$ matrix defined by:
\begin{equation}
\bDelta = \esp_{p_\text{ref}(\by)}\{ \bkappa_{\bomega}(\by) \bkappa_{\bomega}(\by)^\top \otimes \bkappa_{\bomega}(\by) \label{Delta}
\}
\end{equation}
The expression of $\bDelta$ is given in  \ref{Computation of Gamma and Delta}. 

\item Application of the MIA implies $\bN=\bzero$. 
\end{itemize}

The variance of the test statistics $g_t$ in \eqref{test_stat} can be calculated using the independence assumption required previously for the computation of its mean. In particular:
\begin{align}
\label{variance_detector}
     \var \{g_t\} &= \esp \{g(\by_{t})^2\} - \esp \{g(\by_{t})\}^2\nonumber\\
         & = \frac{1}{N_{test}}\left(\tr(\bH \bC_{\btheta,t}) - (\bh^\top \bm_{\btheta,t} )^2\right)
\end{align}

The mean term $\bm_{\btheta,t}$ in \eqref{eq:R'}, \eqref{variance_detector} equals zero when $\btheta_0 = \bzero$ and can be neglected for large values of $t$ according to the mean analysis in Section \ref{mean analysis}. As a consequence, setting $\bZ= \bzero$, vectorizing \eqref{iterC_H1} and using standard results on Kronecker product leads to the following proposition.

\begin{proposition} \label{proposition1}
Under the null hypothesis, neglecting $\bm_{\btheta,t}$ and assuming the MIA holds, $\bc_{\btheta,t} = \vect(\bC_{\btheta,t})$ verifies:
\begin{equation} 
	\bc_{\btheta,t+1} = \bS \bc_{\btheta,t} + \mu^2 \vect(\bQ) \label{rec vecC}
\end{equation}
with:
$$
\bS = (1-\mu\nu)^2\bI +\frac{\mu^2}{N_\text{ref}}(\bGamma +(N_\text{ref}-1)\bH\otimes \bH) -\mu(1-\mu\nu)(\bH \oplus \bH)
$$
where $\bH \oplus \bH = \bH\otimes \bI + \bI\otimes \bH$. 
The variance of the test statistics  \eqref{test_stat} is given~by:
\begin{equation}
\var \{g_t\} = \frac{1}{N_{test}}\tr(\bH \bC_{\btheta,t}) \label{variance_detector_2}
\end{equation}
\end{proposition}

The stability of matrix $\bS$ then ensures the mean-square stability of the algorithm. If the algorithm is mean-square stable, then,  $\bc_{\btheta,t}$ converges to:
\begin{equation}
\label{steady_state}
	\bc_{\btheta,\infty} = \mu^2 (\bI - \bS)^{-1}\vect(\bQ)
\end{equation}
The asymptotic variance of the test statistics directly derives from this result.
Assuming a small step size $\mu$ we have:
$$
\bS = \bI- 2\mu\nu \bI - \mu \bH \oplus \bH + o(\mu) 
$$
Replacing in \eqref{variance_detector_2} and using classical properties of Kronecker products, the asymptotic variance simplifies to:
\begin{align}
\var\{g_\infty\} &=  \frac{\mu}{N_\text{test}}\tr\left(\bH \vect^{-1}\left( (2\nu \bI + \bH \oplus \bH)^{-1}\vect(\bQ) \right)\right) + o(\mu) \label{lyap}\\
&=\frac{\mu}{N_\text{test}} \vect(\bH) (2\nu \bI + \bH \oplus \bH)^{-1}\vect(\bQ) + o(\mu) \label{var_H0_approx}
\end{align}
Note that the rightmost term in \eqref{lyap} can be efficiently computed as the solution of the Lyapunov equation
$(\nu \bI+ \bH ) \bX + \bX (\nu \bI + \bH)=\bQ$
, see \cite[Proposition 7.2.4]{Bernstein2009}.

\subsection{Stochastic behavior analysis in the presence of a change-point}
\label{sec:Stochastic behavior analysis under H1}

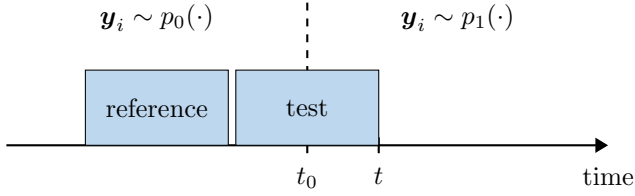
\begin{figure}[!t]
\begin{center}
\begin{tikzpicture}

\draw[thick, -Triangle] (4cm,0) -- (12cm,0)node[right]{};  
\draw (12,0) node[below=5pt]{time};
\draw (8,0) node[below=5pt]{$t_0$};
\draw[thick] (8.95,-3pt) -- (8.95,3pt);
\draw (8.95,0) node[below=5pt]{$t$};
\draw[thick, dashed] (8,-3pt) -- (8, 2cm);

\node[draw, fill=mycolor, minimum width=1.9cm, minimum height=1cm] at (6,0.5) {reference};

\node[draw, fill=mycolor, minimum width=1.9cm, minimum height=1cm] at (8,0.5) {test};

\node[draw=none,fill=none] at (6,1.7) {$\boldsymbol{y}_i\sim p_0(\cdot)$};

\node[draw=none,fill=none] at (10,1.7) {$\boldsymbol{y}_i\sim p_1(\cdot)$};
\end{tikzpicture}
\end{center}
\vspace{-5mm}
\caption{\footnotesize{An illustration of CPD procedure when the test interval contains samples driven by $p_0(\cdot)$ and $p_1(\cdot)$.}  \label{fig_1}}
\end{figure}

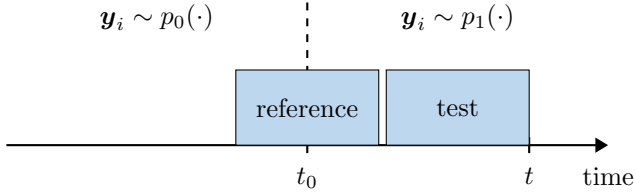
\begin{figure}
\begin{center}
\begin{tikzpicture}
\draw[thick, -Triangle] (4cm,0) -- (12cm,0)node[right]{};  
\draw (12,0) node[below=5pt]{time};
\draw[thick, dashed] (8,-3pt) -- (8, 2cm);
\draw (8,0) node[below=5pt]{$t_0$};
\draw[thick] (10.95,-3pt) -- (10.95,3pt);
\draw (10.95,0) node[below=5pt]{$t$};

\node[draw, fill=mycolor, minimum width=1.9cm, minimum height=1cm] at (8,0.5) {reference};

\node[draw, fill=mycolor, minimum width=1.9cm, minimum height=1cm] at (10,0.5) {test};

\node[draw=none,fill=none] at (6,1.7) {$\boldsymbol{y}_i\sim p_0(\cdot)$};

\node[draw=none,fill=none] at (10,1.7) {$\boldsymbol{y}_i\sim p_1(\cdot)$};
\end{tikzpicture}
\end{center}
\vspace{-5mm}
\caption{\footnotesize{An illustration of CPD procedure when the reference interval contains samples driven by $p_0(\cdot)$ and $p_1(\cdot)$.}  \label{fig_2}}
\end{figure}

Under the assumption of the presence of a single change-point $t_0$, the analysis is conducted by comparing each time instant $t$ to $t_0$, as~$\bsh^\text{ref}_t$, $\bsh^\text{test}_t$ and $\bsH^\text{ref}_t$ defined by \eqref{esp_ht}--\eqref{esp_H} depend on time $t$. We assume that input data $\by_i$ are i.i.d with $\by_i\sim p_0(\cdot)$ before the change, and i.i.d with $\by_i\sim p_1(\cdot)$ after the change. 
\begin{itemize}
    \item If $t<t_0$:  $\bsh^\text{test}_t = \bsh^\text{ref}_t = \bh_0$ and $\bsH^\text{ref}_t = \bH_0$.

	\item If $t_0 \leq t \leq t_0 + N_{\text{test}} - 1$: the test interval contains samples from both distributions; see figure~\ref{fig_1}. According to~\eqref{esp_ht}: 
    \begin{align*} 
	\bsh^\text{test}_t  
    &= \frac{1}{N_{\text{test}}} \sum^t_{i=t-(N_\text{test}-1)} \esp \{\bkappa_{\bomega} (\by_i)\}\\
    &= \frac{1}{N_{\text{test}}} ~\big(n_{0} ~ \bh_0 + n_{1}~\bh_1\big)
	\end{align*}
    where  $n_{1} = t - t_0 + 1$, and $n_{0} =  N_{\text{test}} - n_{1}$.
    In that case: $\bsh^\text{ref}_t = \bh_0$ and $\bsH^\text{ref}_t = \bH_0$. 
    
    \item If $t_0 + N_{\text{test}} \leq t \leq t_0 + N_{\text{test}} + N_{\text{ref}} - 1 $: the reference interval contains samples from both distributions; see figure~\ref{fig_2}. In the same way we find:
    \begin{align*} 
	&\bsh^\text{ref}_t  =  \frac{1}{N_{\text{ref}}}~ (n_0'~ \bh_0 + n_1'~\bh_1)\\
    &\bsH^\text{ref}_t  =  \frac{1}{N_{\text{ref}}}~ (n_0' ~ \bH_0 + n_1'~\bH_1)
	\end{align*}
    where: $n_{1}'=t-(t_0+N_{\text{test}})+1$, and $n_{0}'=N_{\text{ref}}-n_{1}'$.
    \item If $t \geq t_0 + N_{\text{ref}} + N_{\text{test}}$,  $\bsh^\text{test}_t = \bsh^\text{ref}_t = \bh_1$ and $\bsH^\text{ref}_t= \bH_1$.
\end{itemize}
$\bh_0$, $\bh_1$, $\bH_0$ and $\bH_1$ can be computed using the expressions in  \ref{Computation of H and h} when $p_0$ is $\mathcal{N}(\bmu_0, \bR_0)$ and $p_1$ is $\mathcal{N}(\bmu_1, \bR_1)$.

\subsubsection{Mean analysis}

A recursive model of $\bm_{\btheta, t}$ can be obtained by replacing $\bsh^\text{ref}_t$, $\bsh^\text{test}_t$, and  $\bsH^\text{ref}_t$ by their expressions over time in \eqref{weight_error_bias}.
\begin{itemize}
    \item If $t < t_0$:
	\begin{equation} 
	\bm_{\btheta, t+1} =  \big[\bI - \mu(\bH_{0}+ \nu \bI) \big]\bm_{\btheta, t} \label{mean_theta_H1_1}
	\end{equation}
	
	\item If $t_0 \leq t \leq t_0 + N_{\text{test}} - 1$:
	\begin{equation} 
	\bm_{\btheta, t+1} =  \big[\bI - \mu(\bH_{0}+ \nu \bI) \big]~\bm_{\btheta, t} + \mu ~ \frac{n_{1}}{N_{\text{test}}} ~(\bh_1 - \bh_0) \label{mean_theta_H1_2}
	\end{equation}
    
    \item If $ t_0 + N_{\text{test}} \leq t \leq t_0 + N_{\text{test}} + N_{\text{ref}} - 1 $:	\begin{align} 
	 \bm_{\btheta, t+1} &=  \big[\bI - \mu~(\frac{n_{0}'}{N_{\text{ref}}}~ \bH_0 + \frac{n_{1}'}{N_{\text{ref}}}~\bH_1+ \nu \bI) \big]~\bm_{\btheta, t} \nonumber\\
	 &+ \mu ~\frac{n_{0}'}{N_{\text{ref}}}~ (\bh_1 - \bh_0)
	 \label{mean_theta_H1_3}
	\end{align} 
    
    \item If $t \geq t_0 + N_{\text{ref}} + N_{\text{test}}$:
	\begin{equation}
\bm_{\btheta, t+1} =  \big[\bI - \mu(\bH_1+ \nu \bI) \big]\bm_{\btheta, t}\label{mean_theta_H1_4}
\end{equation}
and the mean stability of the algorithm is ensured by using a step size $\mu$ that satisfies:	
    \begin{equation*}
	\mu< \frac{2}{\zeta_{\max}\{\bH_1+\nu\bI\}}
    \end{equation*}

\end{itemize}

The mean of the test statistics~\eqref{test_stat} is, in the presence of a change-point, given by:
\begin{equation}
\esp \{g_{t}\} = 
\left\{
\begin{aligned}
&\bh_0^\top~\bm_{\btheta, t} &&  t<t_0 \\
&\frac{1}{N_\text{test}}(n_1\bh_1+n_0\bh_0)^\top\bm_{\btheta, t} &&  t_0 \leq t<t_0+N_{\text{test}}\\
&\bh_1^\top~\bm_{\btheta, t} &&  t\geq t_0  + N_{\text{test}}
\end{aligned}
\right. \label{mean_detector_H1}
\end{equation}

\subsubsection{Mean squared analysis}
The first step consists of the computation of the matrices $\bT$, $\bQ$, $\bZ$ and $\bN$ in the presence of a change point.
\begin{itemize}
\item Following the same steps as in~\eqref{eq:T}, we find:
\begin{equation}
\bT = \frac{1}{N_\text{ref}}~ \Big( \vect^{-1}(\bGamma \bc_{\btheta,t}) + (N_\text{ref}-1)~  \bsH^\text{ref}_{t+1} \bC_{\btheta,t} \bsH^\text{ref}_{t+1}\Big) \label{eq:T_H1}    
\end{equation}
where $\bGamma$ is defined in \eqref{Gamma} and depends on $t$.

\item $\bQ$ can be decomposed as:
\begin{equation}
	\bQ = \bQ_1 + \bQ_2 - (\bQ_3 + \bQ_3^\top)
\end{equation}
where: 
\begin{align}
\label{Q1}
\bQ_{1} &= \esp \{\bh_{t+1}^\text{test}~{\bh_{t+1}^\text{test}}^\top\}
\end{align} 
Substituting~\eqref{h_test} into \eqref{Q1} and expanding the expression we get:
\begin{equation*}
\bQ_{1}= \frac{1}{N_{\text{test}}}~\bsH^\text{test}_{t+1} + (1- \frac{1}{N_{\text{test}}})~ \bsh^\text{test}_{t+1} (\bsh^\text{test}_{t+1})^\top
\end{equation*}  
where: $$\bsH^\text{test}_t = \esp_{p_\text{test}(\by)}\{\bH_{t}^\text{test}\}\label{esp_Ht}$$

In the same way, we find:
\begin{align*}
\bQ_{2} &=\esp \{\bh_{t+1}^\text{ref}~{\bh_{t+1}^\text{ref}}^\top\}\\
&= \frac{1}{N_{\text{ref}}}~\bsH^\text{ref}_{t+1} + (1- \frac{1}{N_{\text{ref}}})~ \bsh^\text{ref}_{t+1} (\bsh^\text{ref}_{t+1})^\top
\end{align*}
and since the samples $\by_i$ in the reference and test intervals are independent,
\begin{align*}
\bQ_{3} &= \esp \{ \bh^\text{test} {\bh^\text{ref}}^\top\} \\
&=  \bsh^\text{test}_{t+1} (\bsh^\text{ref}_{t+1})^\top
\end{align*}

\item The matrix $\bZ$ can be expanded as:
\begin{equation}
\bZ = \esp\{ \bh^\text{ref}_{t+1} \btheta_t^\top \bH^\text{ref}_{t+1}\} - \esp\{ \bh^\text{test}_{t+1} \btheta_t^\top \bH^\text{ref}_{t+1}\}\label{Z}
\end{equation}
The first expectation term in \eqref{Z} can be computed using the MIA and the vectorization operator:
\begin{align}
\bZ &= \frac{1}{N_{\text{ref}}}~ \big( \vect^{-1}(\bDelta \bm_{\btheta, t}) + (N_{\text{ref}} - 1)~\bsh^\text{ref}_{t+1}  \bm_{\btheta, t}^\top \bsH^\text{ref}_{t+1} \big) \nonumber\\ &-\bsh^\text{test}_{t+1} \bm_{\btheta, t}^\top \bsH^\text{ref}_{t+1} 
\end{align}
where $\bDelta$ is defined in \eqref{Delta} and depends on $t$.

\item Using the MIA: 
$$\bN=(\bsh^\text{test}_{t+1} -\bsh^\text{ref}_{t+1})~\bm_{\btheta,t}^\top$$

\end{itemize}

The last step consists in replacing the expressions of $\bsh^\text{ref}_t$, $\bsh^\text{test}_t$, and  $\bsH^\text{ref}_t$ as a function of $t$ in $\bT$, $\bQ$, $\bZ$ and $\bN$. 
We will denote by $\bGamma_0$ (resp. $\bGamma_1$) and  $\bDelta_0$ (resp. $\bDelta_1$) matrices
$\bGamma$ and $\bDelta$ in (\ref{Gamma},\ref{Delta})  computed for $\by_i \sim p_0(\cdot)$ 
(resp. $\by_i \sim p_1(\cdot)$), see \ref{Computation of Gamma and Delta}. This leads to the following proposition.

\begin{proposition} \label{proposition2}
Under the assumption of a change-point $t_0$ and assuming that the MIA holds, $\bC_{\btheta,t}$ is given by \eqref{iterC_H1} where:
\begin{itemize}
    \item If $t<t_0$ :
    \begin{align*}
    \bT &= \frac{1}{N_\text{ref}}~ \Big(\vect^{-1}(\bGamma_0 \bc_{\btheta,t}) + (N_\text{ref} -1)~ \bH_{0} \bC_{\btheta,t} \bH_{0}\Big) \\
    \bQ &= \frac{N_\text{ref}+ N_\text{test}}{N_\text{ref}~N_\text{test}}~( \bH_0 -  \bh_0 \bh_0^\top ) \\
    \bZ &=  \frac{1}{N_{\text{ref}}}~ \big(  \vect^{-1}(\bDelta_0 \bm_{\btheta,t}) - \bh_{0} \bm_{\btheta,t}^\top \bH_{0} \big)\\
    \bN &= \bzero
    \end{align*}
 
    \item If $t_0 \leq t \leq t_0 + N_\text{test}-1$:
    \begin{align*}
    \bT &= \frac{1}{N_\text{ref}}~ \Big(\vect^{-1}(\bGamma_0 \bc_{\btheta,t}) + (N_\text{ref} -1) ~\bH_{0} \bC_{\btheta,t}  \bH_{0}\Big) \\
    \bQ &= \big(\frac{n_{0}}{N_\text{test}^2} + \frac{1}{N_\text{ref}}\big)\bH_0 + \frac{n_{1}}{N_\text{test}^2}\bH_1 + \frac{n_{1}(n_{1}-1)}{N_\text{test}^2}\bh_1 \bh_1^\top \\
    &+ \Big(\frac{n_{0}\big(n_{0}-1\big)}{N_\text{test}^2} - \frac{2n_{0}}{N_\text{test}} - \frac{1}{N_\text{ref}}+ 1 \Big) ~\bh_0\bh_0^\top \\
    &+ n_{1} ~\big(\frac{n_{0}}{N_\text{test}^2}-\frac{1}{N_\text{test}}\big) (\bh_0 \bh_1^\top + \bh_1 \bh_0^\top)\\
    \bZ &= \big(1 \!-\! \frac{1}{N_\text{ref}} \!-\! \frac{n_{0}}{N_\text{test}}\big) ~\bh_0 \bm_{\btheta, t}^\top \bH_0 - \frac{n_{1}}{N_\text{test}} ~\bh_1 \bm_{\btheta, t}^\top \bH_0   \\
    &+\frac{1}{N_\text{ref}}~\vect^{-1}(\bDelta_0 \bm_{\btheta, t})\\
    \bN &= \frac{n_1}{N_{\text{test}}}~(\bh_1 - \bh_0)~\bm_{\btheta, t}^\top
\end{align*}
     \item If $t_0 + N_\text{test} \leq t \leq t_0 + N_\text{test}+ N_\text{ref}-1$
     \begin{align*}
	   \bT &= \frac{1}{N_\text{ref}^2} ~\Big(n_{0}'~ \vect^{-1}(\bGamma_0 \bc_{\btheta,t})+ n_{1}'~\vect^{-1}(\bGamma_1 \bc_{\btheta,t}) \\
	   &+n_{0}'(n_{0}'- 1)\bH_0 \bC_{\btheta,t} \bH_0 + n_{1}'(n_{1}' - 1)\bH_1 \bC_{\btheta,t} \bH_1 \\
	    &+ n_{0}'n_{1}'~(\bH_0 \bC_{\btheta,t} \bH_1 + \bH_1 \bC_{\btheta,t} \bH_0) \Big) \\
	  \bQ &= \frac{n_{0}'}{N_{\text{ref}}^2}~\bH_0 + \big( \frac{n_{1}'}{N_{\text{ref}}^2} + \frac{1}{N_{\text{test}}} \big)~\bH_1 + \frac{n_{0}'(n_{0}'-1)}{N_{\text{ref}}^2}~\bh_0\bh_0^\top \\
	  &+ \Big(\frac{n_{1}'(n_{1}'-1)}{N_{\text{ref}}^2}-\frac{2n_{1}'}{N_{\text{ref}}}-\frac{1}{N_{\text{test}}}+1\Big)~\bh_1\bh_1^\top \\
	  &+ n_{0}'~(\frac{n_{1}'}{N_{\text{ref}}^2} - \frac{1}{N_{\text{ref}}})~(\bh_0\bh_1^\top + \bh_1\bh_0^\top)\\
	  \bZ &= \frac{1}{N_\text{ref}^2}~\Big( n_{0}'\vect^{-1}(\bDelta_0 \bm_{\btheta,t})+ n_{1}'\vect^{-1}(\bDelta_1 \bm_{\btheta,t}) \\
	  &+ n_{0}'(n_{0}' - 1)~\bh_0  \bm_{\btheta,t}^\top \bH_0 + n_{0}' n_{1}' ~\bh_0 \bm_{\btheta,t}^\top \bH_1 \\
	  &- n_{0}'^2~ \bh_1  \bm_{\btheta,t}^\top \bH_0 - n_{1}'(n_{0}'+1)~ \bh_1 \bm_{\btheta, t}^\top \bH_1\Big)   \\
	    \bN &= \frac{n_{0}'}{N_{\text{ref}}}~(\bh_1 - \bh_0)~\bm_{\btheta, t}^\top
\end{align*}

    \item If $t \geq t_0 + N_\text{test} +  N_\text{ref}$
    \begin{align*}
    \bT &= \frac{1}{N_\text{ref}} ~\Big(\vect^{-1}(\bGamma_1 \bc_{\btheta,t}) + (N_\text{ref}-1)~ \bH_{1} \bC_{\btheta,t} \bH_{1}^{\top}\Big) \\
    \bQ &= \frac{N_\text{ref}+ N_\text{test}}{N_\text{ref}~N_\text{test}} ~( \bH_1 -  \bh_1 \bh_1^\top ) \\
    \bZ &=  \frac{1}{N_\text{ref}} ~\big(\vect^{-1}(\bDelta_1 \bm_{\btheta,t}) - \bh_{1} \bm_{\btheta,t}^\top \bH_{1} \big)\\
    \bN &= \bzero
    \end{align*}
\end{itemize}

The variance of the test statistics  \eqref{test_stat} is given by:
\begin{equation}
\label{variance_detector_H1}
\begin{split}
N_\text{test}&\var\{g_t\}\\&= \left\{
\begin{aligned}
  &\tr(\bH_0 \bC_{\btheta, t})  - (\bh_0^\top \bm_{\btheta, t})^2 &&  t < t_0 \\
  &\tr(\bsH^\text{test}_t \bC_{\btheta, t})  - (\bsh^{\text{test}, \top}_t \bm_{\btheta, t})^2 && t_0 \leq t  < t_0 + N_\text{test} \\ 
  &\tr(\bH_1 \bC_{\btheta, t}) - (\bh_1^\top \bm_{\btheta, t})^2&& t \geq t_0 + N_\text{test}
\end{aligned}
\right.
\end{split}
\end{equation}
\end{proposition}

All these expressions can be further simplified by neglecting $\bm_{\btheta, t}$, specifically when $t<t_0$ if e.g. $\btheta_0=\bzero$ and $t\gg t_0$ assuming mean stability.

\section{Simulation results}
\label{Simulations}

The \texttt{julia} code to reproduce all these experiments will be made available at \url{github.com/andferrari}. 

\subsection{Model validation} \label{Monte Carlo validation}

In this subsection, we present Monte Carlo simulations to illustrate the accuracy of the models derived in Section \ref{Theoretical analysis}. Analytical expressions of the mean and the variance of the detection statistics under the null hypothesis are first considered. The observations $\by_i$ were zero-mean two-dimensional i.i.d Gaussian vectors, with correlation coefficient equal to 0.25, and standard deviation equal to 0.5. Under these assumptions and for a Gaussian reproducing kernel, expressions of $\bh$ and $\bH$ are given in \ref{Computation of H and h}. The algorithm parameters were set as follows: the bandwidth of the Gaussian kernel was $\sigma = 0.25$, the regularization parameter $\nu=10^{-3}$, the step-size $\mu=5.10^{-4}$. The windows lengths were set to $N_\text{ref} = N_\text{test} = 250$, and the $L = 16$ dictionary elements were obtained by sampling the same distribution as $\by_i$.  The results were averaged over 500 Monte Carlo runs.

\begin{figure}
	\begin{center}	

	   \includegraphics[width=.8\columnwidth]{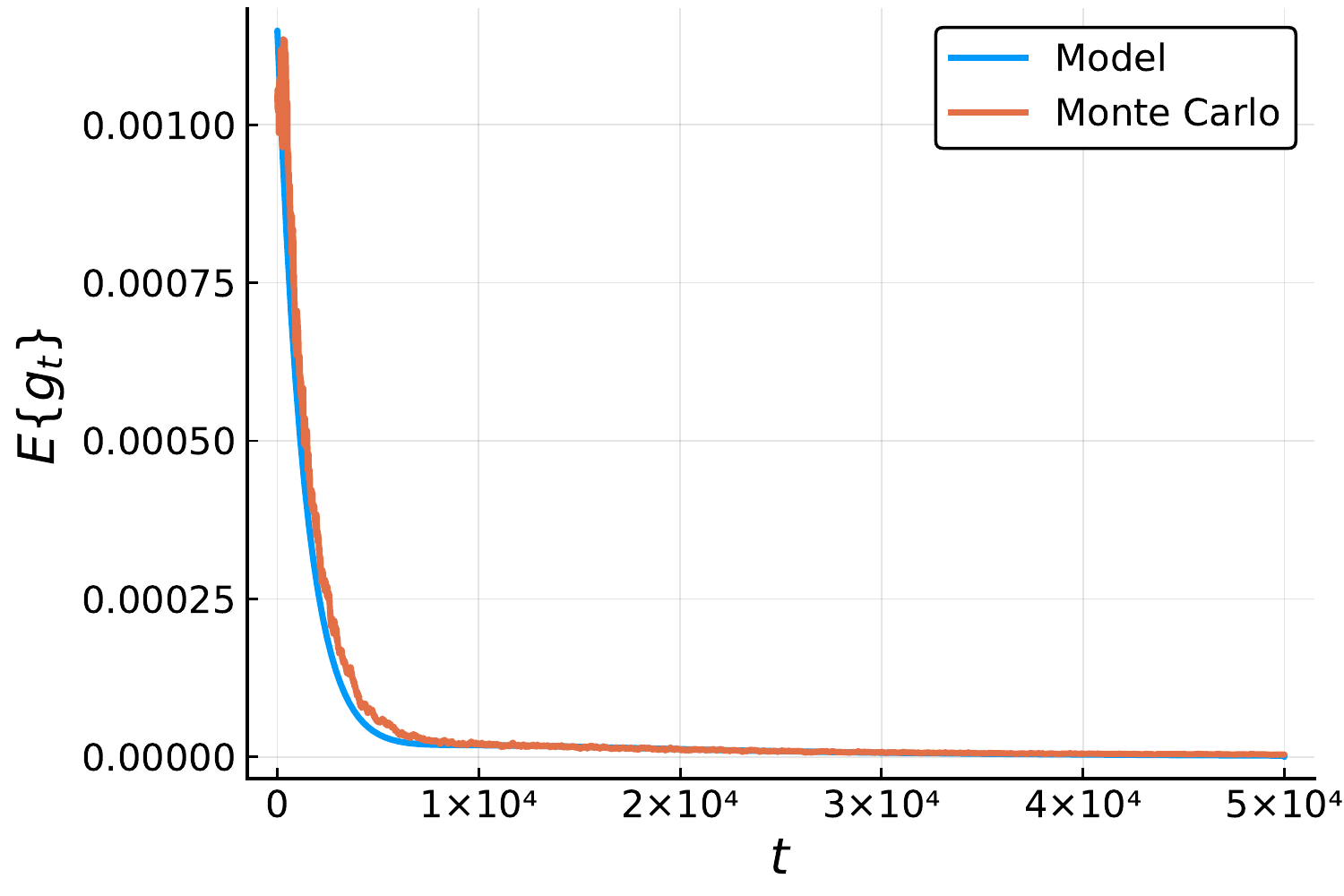}
	   	\end{center}
	\caption{Mean of NOUGAT detection statistics obtained using model \eqref{mean_detector} and Monte Carlo simulations under the null hypothesis.}
\label{mean_stat_H0}
\end{figure}

\begin{figure}

	\begin{center}			
	\includegraphics[width=.8\columnwidth]{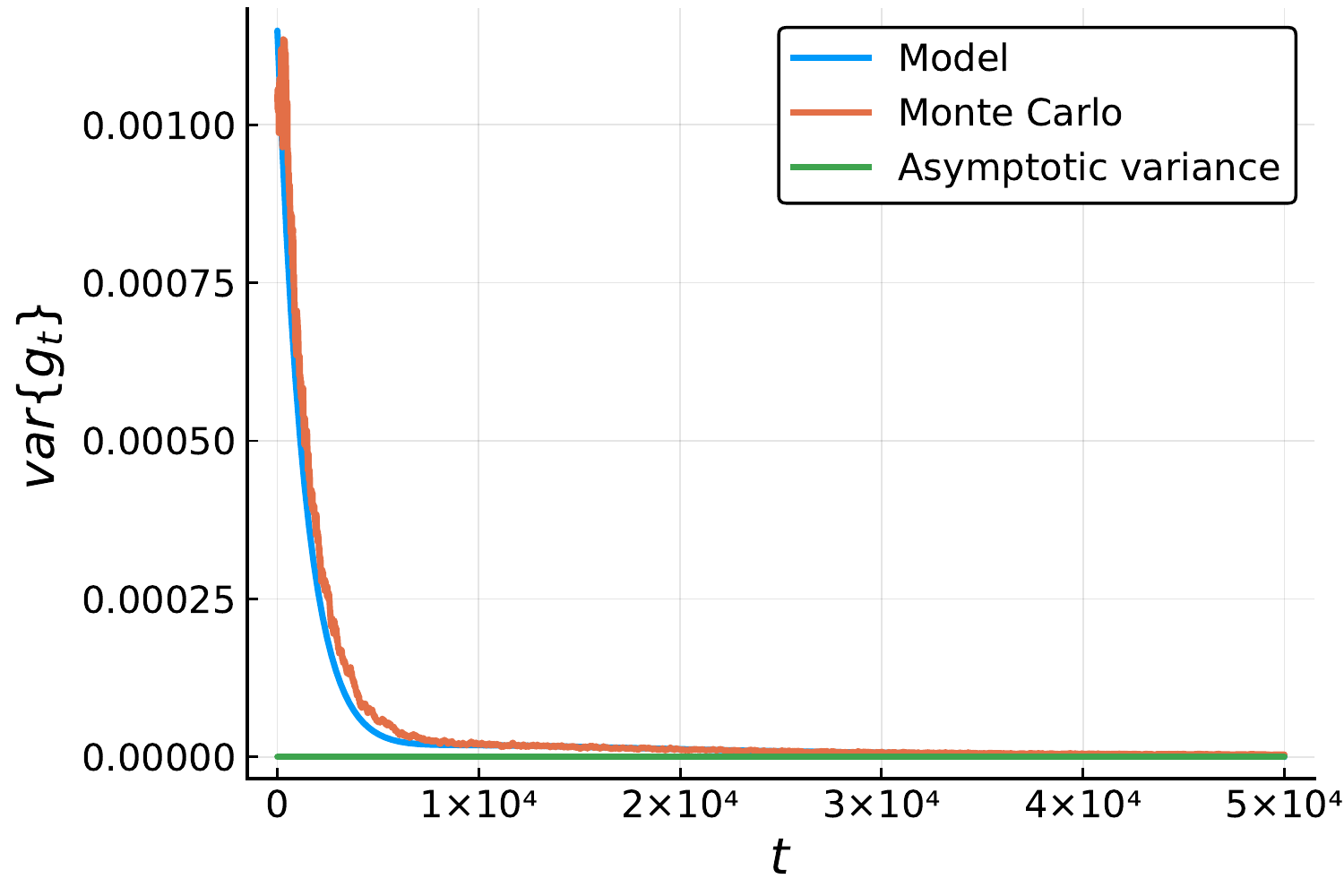}
	\end{center}
		\caption{Variance of NOUGAT detection statistics obtained using model \eqref{variance_detector} and Monte Carlo simulations under the null hypothesis.}
\label{variance_stat_H0}
\end{figure}

Figures~\ref{mean_stat_H0} and~\ref{variance_stat_H0} compare respectively the theoretical models of the mean given by \eqref{weight_mean_underNull}, \eqref{mean_detector} and the variance given in Proposition \ref{proposition1} of the detection statistics, to Monte Carlo simulations. The asymptotic value of the variance computed from \eqref{steady_state} is also reported. The initial weight vector was set to $\btheta_0=(0.3,\,0.3)^\top$. The simulation results clearly show a good accuracy between the models and the actual performance provided by Monte Carlo simulations. These results also confirm the asymptotic unbiasedness of the estimator: $\esp \{g_t\}$ converges to 0 as expected, and validate the assumptions used in the derivations. 

\begin{figure}
	\begin{center}			
	\includegraphics[width=.8\columnwidth]{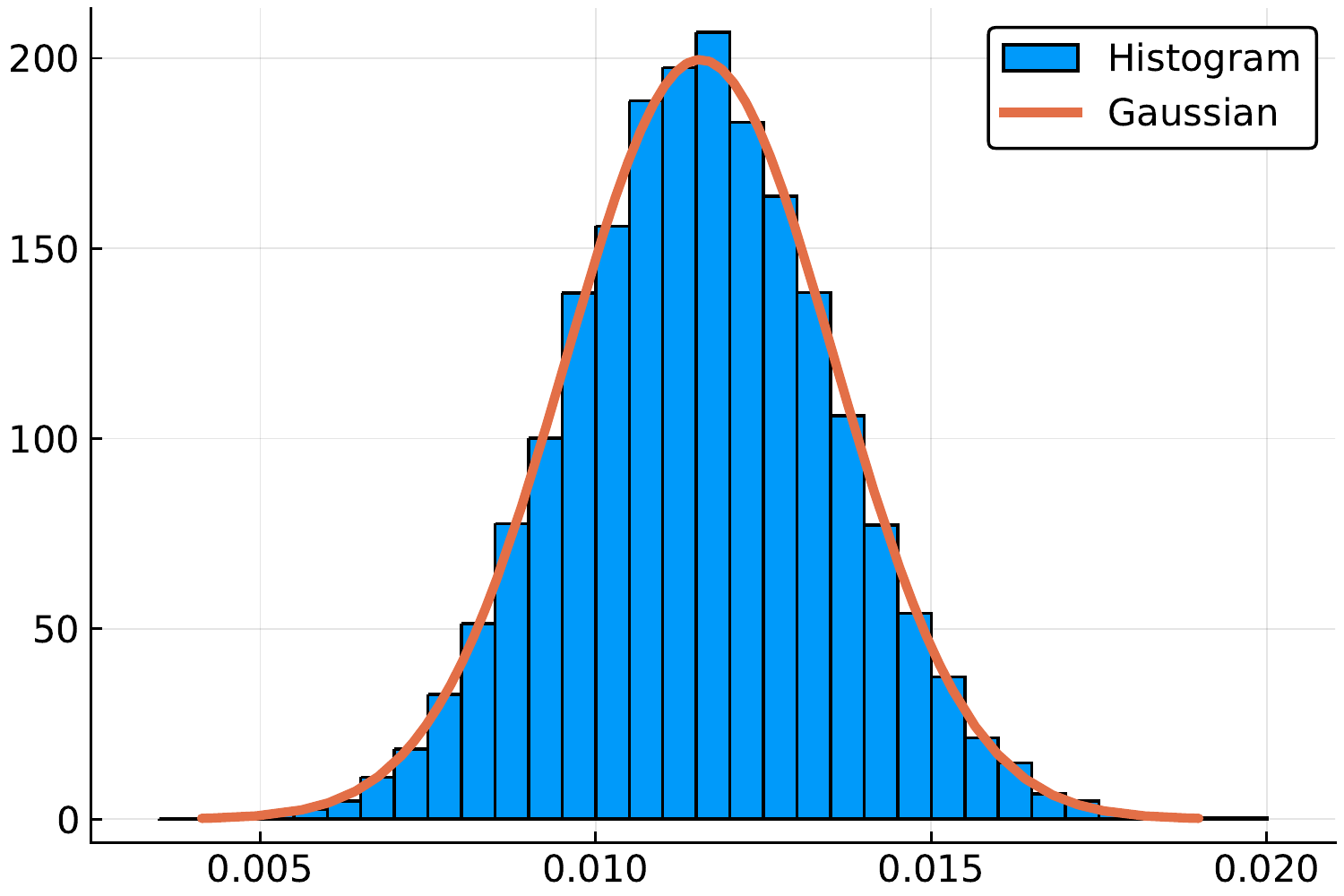}
	\end{center}
	\caption{Histogram of NOUGAT detection statistics under the null hypothesis.  \label{Histogram_H0}}
\end{figure}

\begin{figure}
	\begin{center}			
	\includegraphics[width=.8\columnwidth]{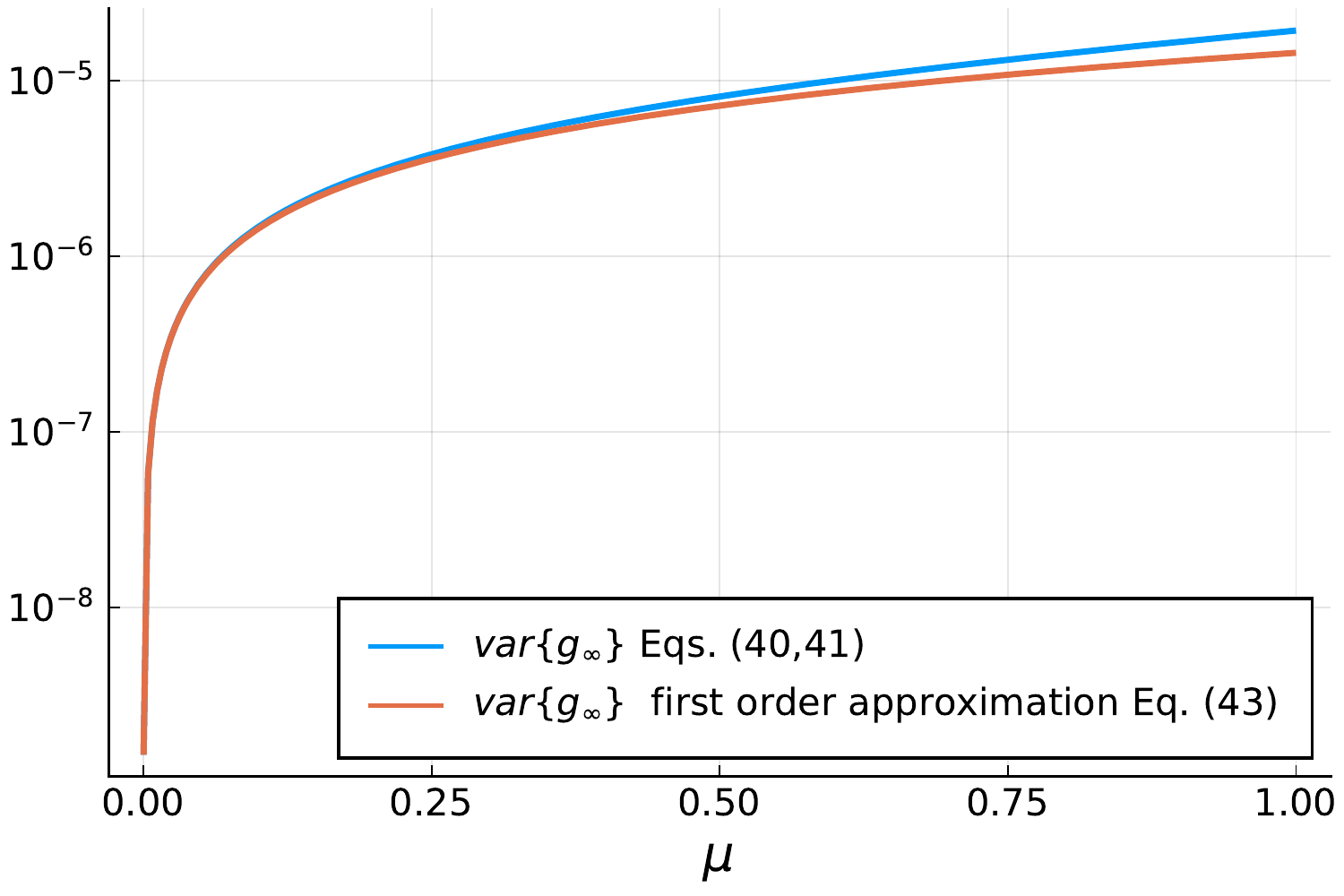}
	\end{center}
	\caption{Asymptotic variance of the test statistic compared to its first order approximation, as a function of the step size $\mu$.  \label{variance_H0}}
\end{figure}

We also provide the histogram of the detection statistics in Figure~\ref{Histogram_H0}. Contrarily to \cite{Ikram2018}, the histogram is very close to its Gaussian approximation as reported in this figure. Note that, as proved above, the mean converges towards zero for larger values of $t$. The accuracy of $g_t$ Gaussian approximation is a central result to set the threshold and guarantee a given false alarm rate using, e.g., the asymptotic expression of \eqref{variance_detector} computed using \eqref{steady_state}. Figures~\ref{variance_H0} compares the asymptotic variance of the test statistic computed using \eqref{variance_detector_2}, \eqref{steady_state} and its first order approximation \eqref{var_H0_approx} when the step size $\mu$ is small ($\mu=5.10^{-4}$ in this section). Note that this expression depends on $\bh$ and $\bH$ which can be computed by Monte Carlo simulations in the  non Gaussian case.

\begin{figure}
	\begin{center}			
\includegraphics[width=.8\columnwidth]{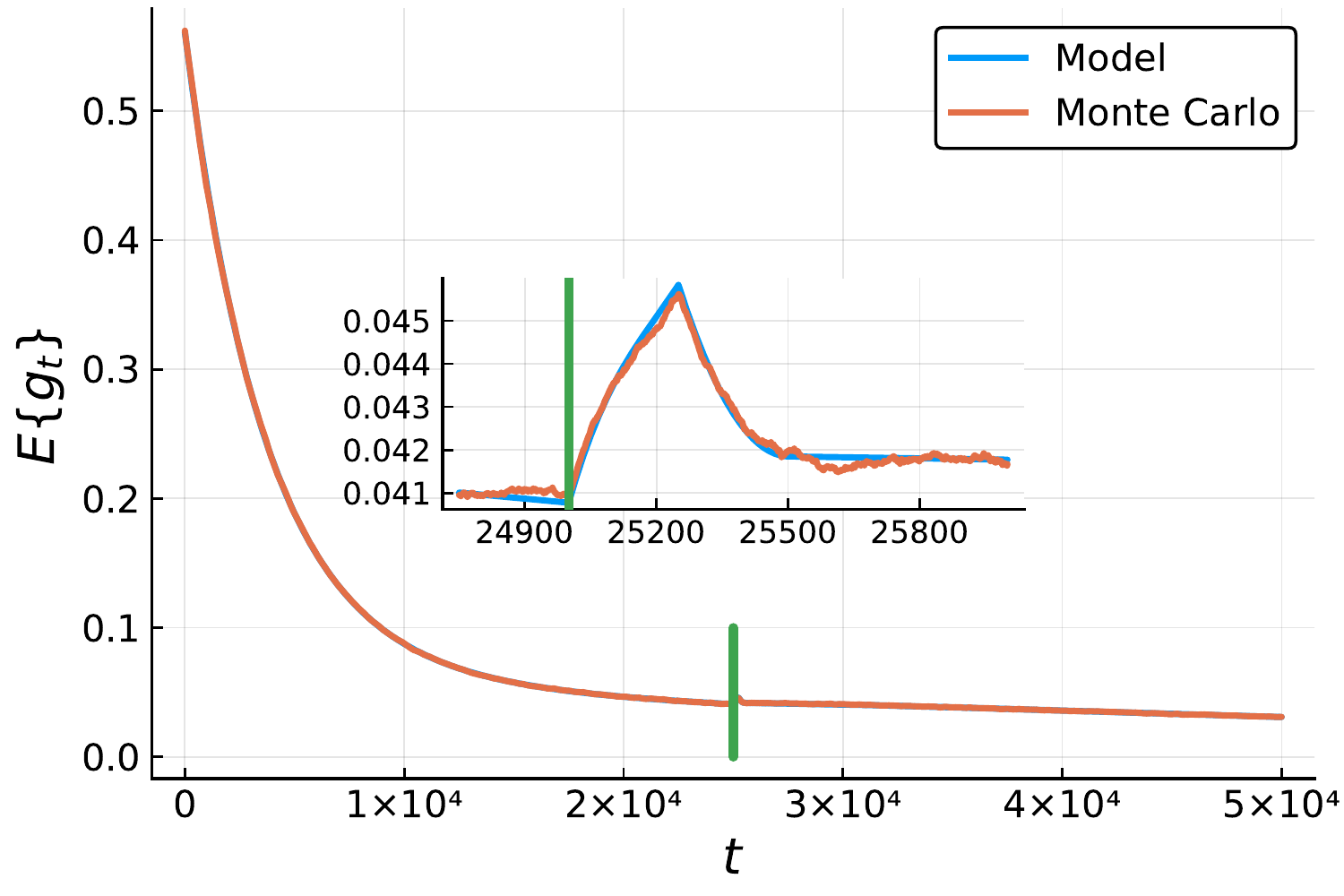}
	\end{center}
	\caption{Mean of NOUGAT detection statistics obtained using the model \eqref{mean_detector_H1} and Monte Carlo simulations. The change is identified by the green line. 
	\label{mean_stat_H1}}
\end{figure}

\begin{figure}
	\begin{center}			
\includegraphics[width=.8\columnwidth]{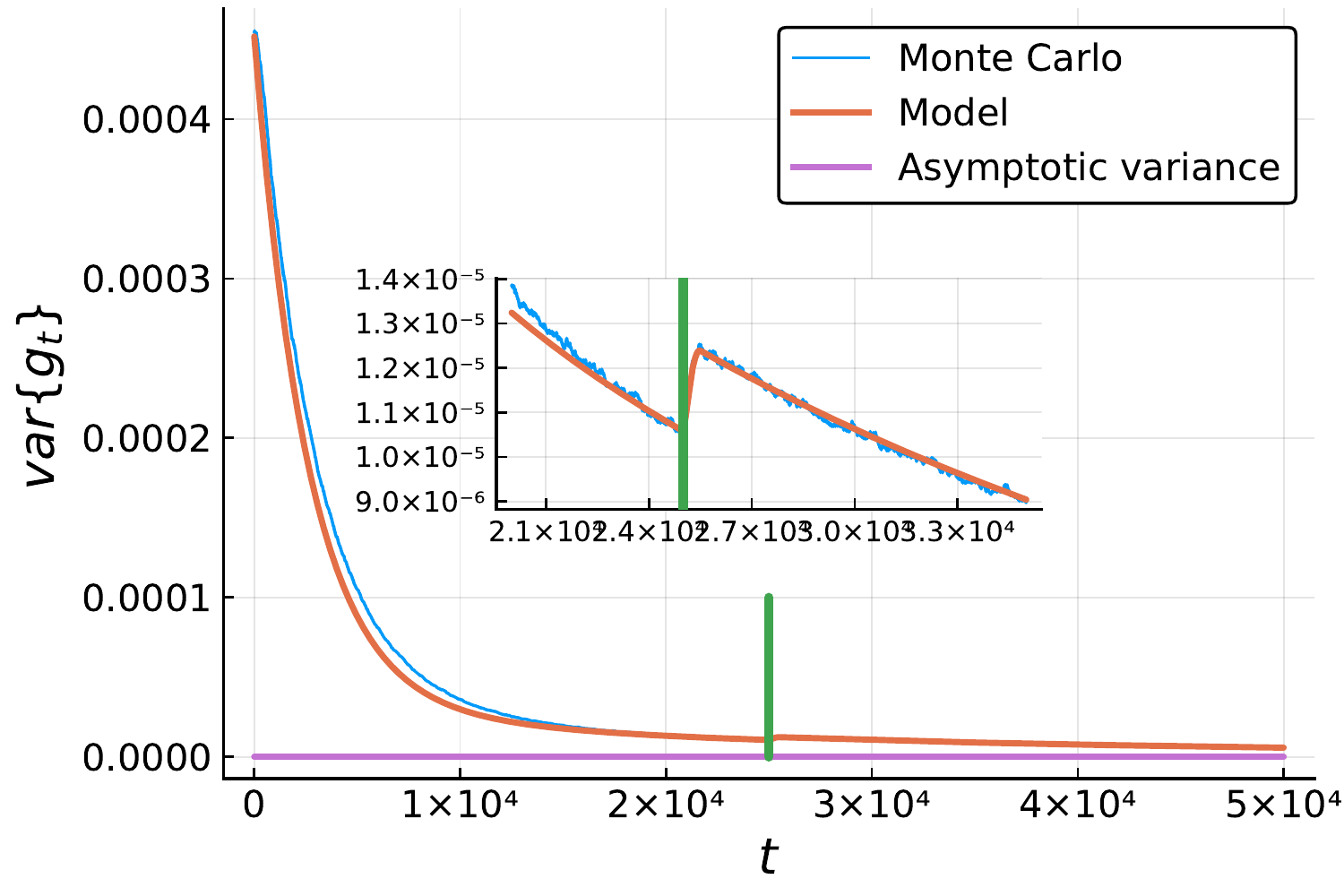}
	\end{center}
	\caption{Variance of NOUGAT detection statistics obtained using model \eqref{variance_detector_H1} and Monte Carlo simulations. The change is identified by the green line. \label{variance_stat_H1}}
\end{figure}

For the second part of the simulations, we inserted a change-point at time instant $t_0 = 25\cdot10^3$ by changing the input vectors correlation coefficient to 0.1 and standard deviation to 0.7. The results for the mean behavior are given in Figure~\ref{mean_stat_H1}, and for the variance in Figure~\ref{variance_stat_H1}.  Both figures clearly show that the theoretical curves provided by~\eqref{mean_theta_H1_1}-\eqref{mean_detector_H1} and Proposition \ref{proposition2} match well the actual performance provided by Monte Carlo simulations, especially in the vicinity of the change point.

\begin{figure}
	\begin{center}			
	\includegraphics[width=.8\columnwidth]{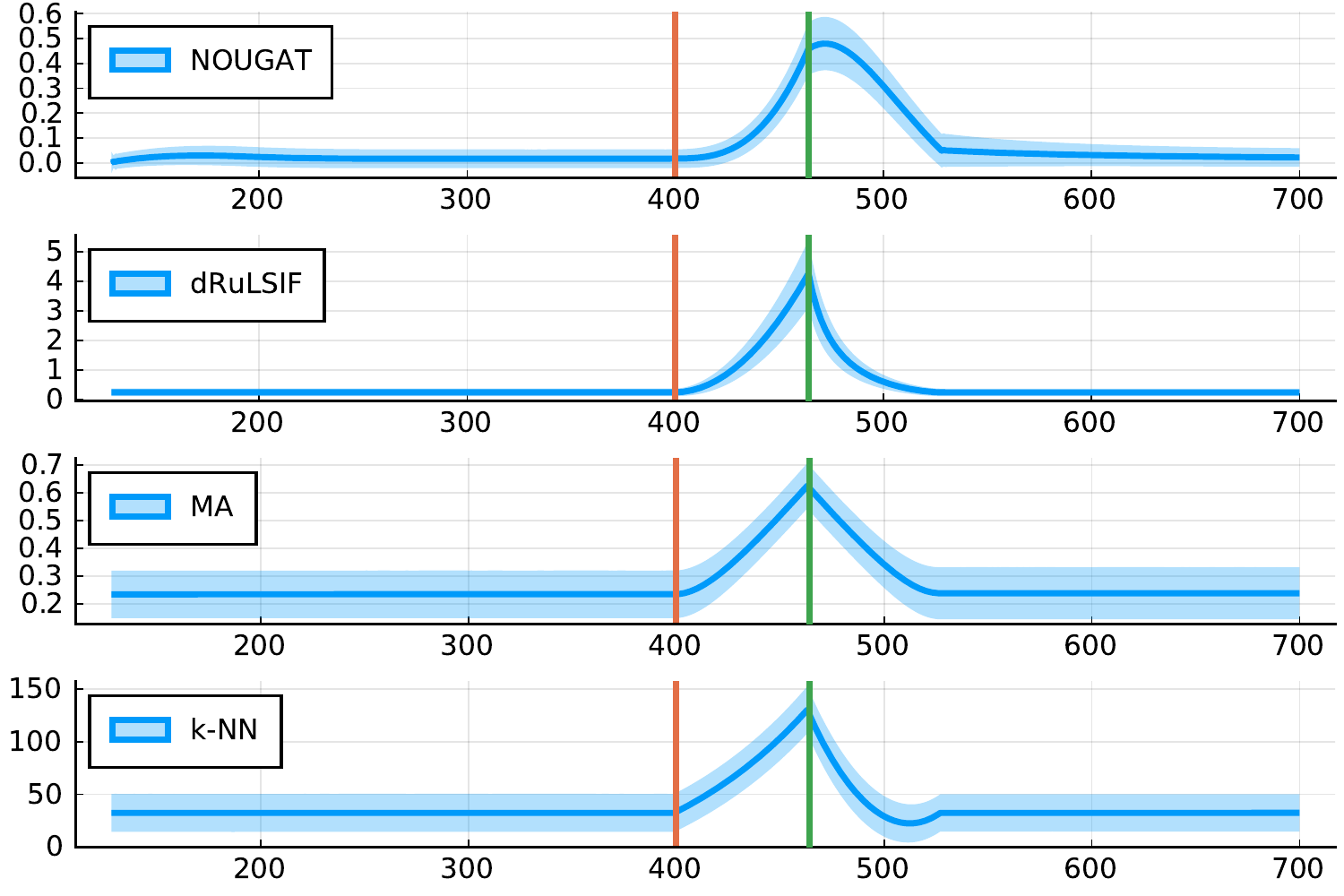}
	\end{center}
	\caption{Mean of the test statistic ($\pm$ standard deviation) 
	  for dRuLSIF, NOUGAT and MA. The change point $t_0$ is located at the red line 
	  and $t_0+N_\text{ref}$ at the green line.}\label{Statistics}
\end{figure}

\subsection{Performances comparison} \label{Performances comparison}

This section aims to compare the performances of 1) dRuLSIF, a debiased version of RuLSIF obtained solving \eqref{opt} at each time instant $t$, 2) NOUGAT, the proposed online version of dRuLSIF, and 3) MA, as defined in \eqref{MA}. Note that all these algorithms share the same memory footprint. 

In order to assess the performance of all these algorithms compared to a non-kernel-based algorithm, we shall now report the detection performance of a nearest-neighbors based CPD algorithm. The algorithm we selected is based on the two-sample test proposed by \cite{schilling_1986, henze_1988}, and recently considered for CPD in~\cite{chen_2019}. At each time instant~$t$, the k-nearest-neighbors algorithm is applied to the samples in the interval $(t-N_\text{ref}-N_\text{test}+1, t)$. The test statistic is related to the number of edges $N_\text{e}$ of the graph that connect observations in the reference window with observations in the test window. Indeed, when these observations are driven by two different distributions, this graph tends to be clustered with respect to the reference and the test window, and $N_\text{e}$ is then ideally close to zero. Correspondingly, the test statistic, denoted as k-NN, is $N_e$ corrected by its mean value under equal distributions hypothesis, see e.g. \cite{chen_2019}. 

The observations $\by_t$ were sampled from a mixture of $n$ $k$-dimensional Gaussian distributions $\mathcal{N}_k(\bm_q,q^{-1}\bC_q)$, with $q=1,\ldots, n$. The weights $\phi_q$ of the mixture model were sampled from a $n$-dimensional Dirichlet distribution of parameter $\alpha$. The means $\bm_q$ were sampled from $\mathcal{N}_k(\bzero,\bI)$ and the covariance matrices $\bC_q$ from a Wishart distribution with scale matrix $\bI$ and $k+2$ degrees of freedom, that is, $\mathcal{W}_k(\bI, k+2)$.

The change point was set to $t_0 = 400$, the number of samples to $n_t = 700$, the dimension of measurements was fixed to $k=6$, the number of mixture components was set to $n=3$ and $\alpha = 5$. All the parameters $(\bm_q, \phi_q, \bC_q)$, with $q=1,\ldots, n$ of the GMM
were resampled at time $t=t_0$.

For all simulations, we considered a Gaussian kernel. Its bandwidth $\sigma$ was set using the median trick, that is, the median of the pairwise distances between samples governed by the same distribution as $\by_t$ under the null hypothesis. A dictionary of $L=80$ elements was designed by sampling the same distribution. For all Monte Carlo simulations, these parameters were kept fixed. For all algorithms, the window lengths were set to $N_\text{ref}=N_\text{test}=64$. The regularization parameter for dRuLSIF and NOUGAT was set to  $\nu = 10^{-2}$ and the step size for NOUGAT was set to $\mu=47\cdot 10^{-3}$. The number of nearest-neighbors of the k-NN was set to 10. This value was obtained experimentally in order to achieve the best performance.

\subsubsection{Detection performance}

Figure~\ref{Statistics} provides the mean $\pm$ standard deviation for the four test statistics, namely, NOUGAT, dRuLSIF, MA and k-NN computed from $10^6$ runs. Note that, contrarily to Figure~\ref{mean_stat_H0},  NOUGAT was initialized with  $\btheta_{-1}=\bzero$ to guarantee unbiasedness under the null hypothesis as shown in Section II.A. 

When comparing the ratio between the peak at $t_0+N_\text{ref}$ (green line) and the noise level for $t<t_0$ (before the red line), Figure~\ref{Statistics} reveals a slight drop in performance of NOUGAT compared to dRuLSIF. A larger loss of performance can be observed with MA and k-NN compared to NOUGAT and dRulSIF. As MA test statistic is the norm of the solution of \eqref{opt} with $\bH_t^\text{ref}=\bI$ and $\nu=0$, it does not take into account correlations in the feature space. In addition, MA does not take advantage of the functional approximation framework as it tests the norm of the parameters vector while NOUGAT approximates the likelihood ratio \eqref{test_stat}. Moreover, we can observe a small detection delay between NOUGAT and the other algorithms. This can be explained by the approximate resolution of \eqref{opt} by a gradient descent step in \eqref{grad_step}. This loss of performance of the online NOUGAT algorithm must be put into perspective, given its much lower computational cost compared to e.g. the offline dRuLSIF algorithm, see Section \ref{Computational cost}. 

To get more insight in the performance of dRuLSIF, NOUGAT and MA, we shall now analyze the Mean Time to False Alarm~(MTFA) and the Mean Time to Detection~(MTD). Both are usual online performance measures~\cite{Gustafssonb}. Let $t_a$ be the time instant of detection and $t_0$ the change point. They are defined as:
\begin{align}
&\text{MTD} = \esp\{t_a-t_0 \,|\, t_a\geq t_0\}\\
&\text{MTFA} = \esp\{t_a \,|\, t_a < t_0\}
\end{align}

\begin{figure}
	\begin{center}			
	\includegraphics[width=.8\columnwidth]{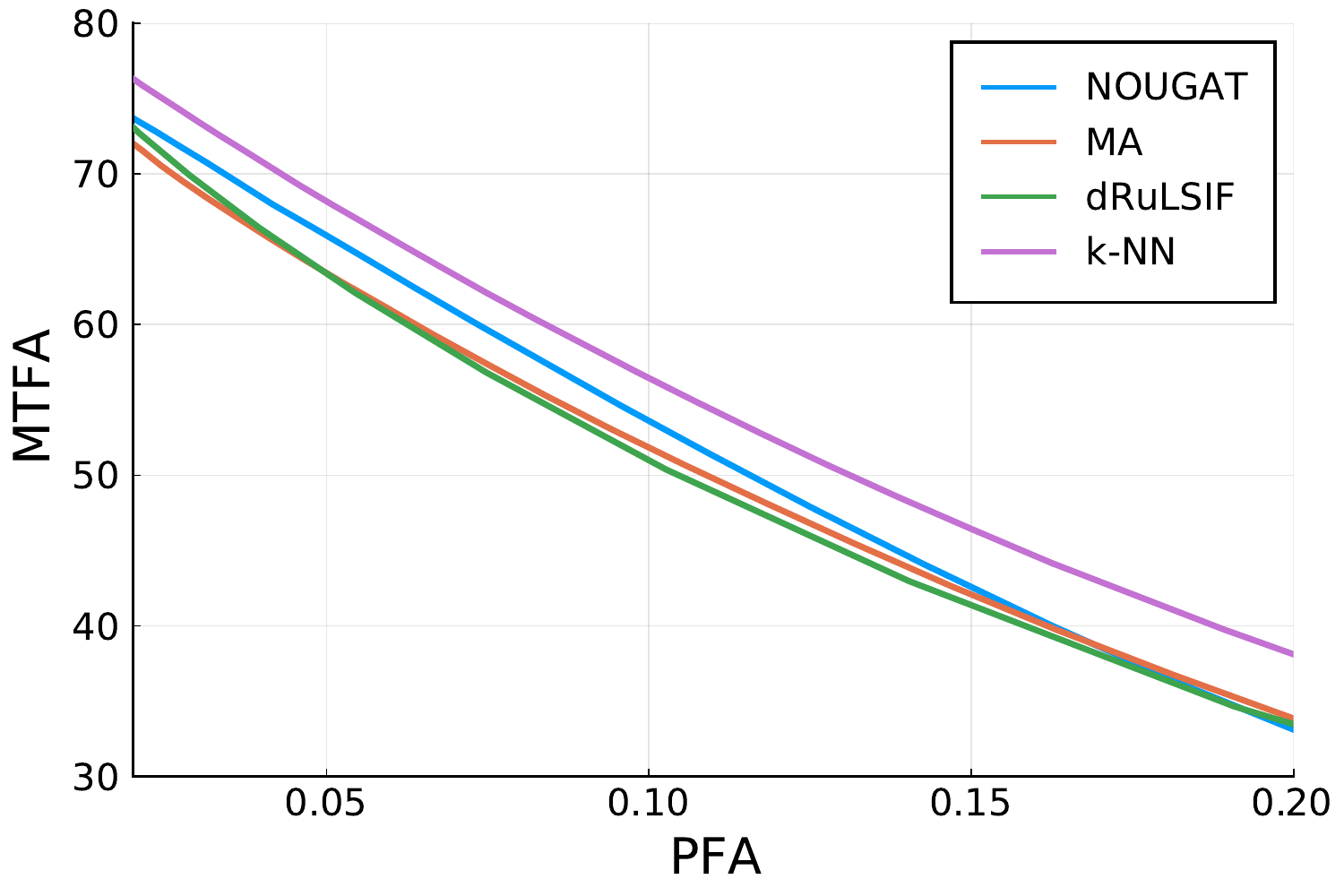}
	\end{center}
	\caption{MTFA as a function of PFA for NOUGAT, MA and dRuLSIF.}\label{MTFA}
\end{figure}

\begin{figure}
	\begin{center}
	\includegraphics[width=.8\columnwidth]{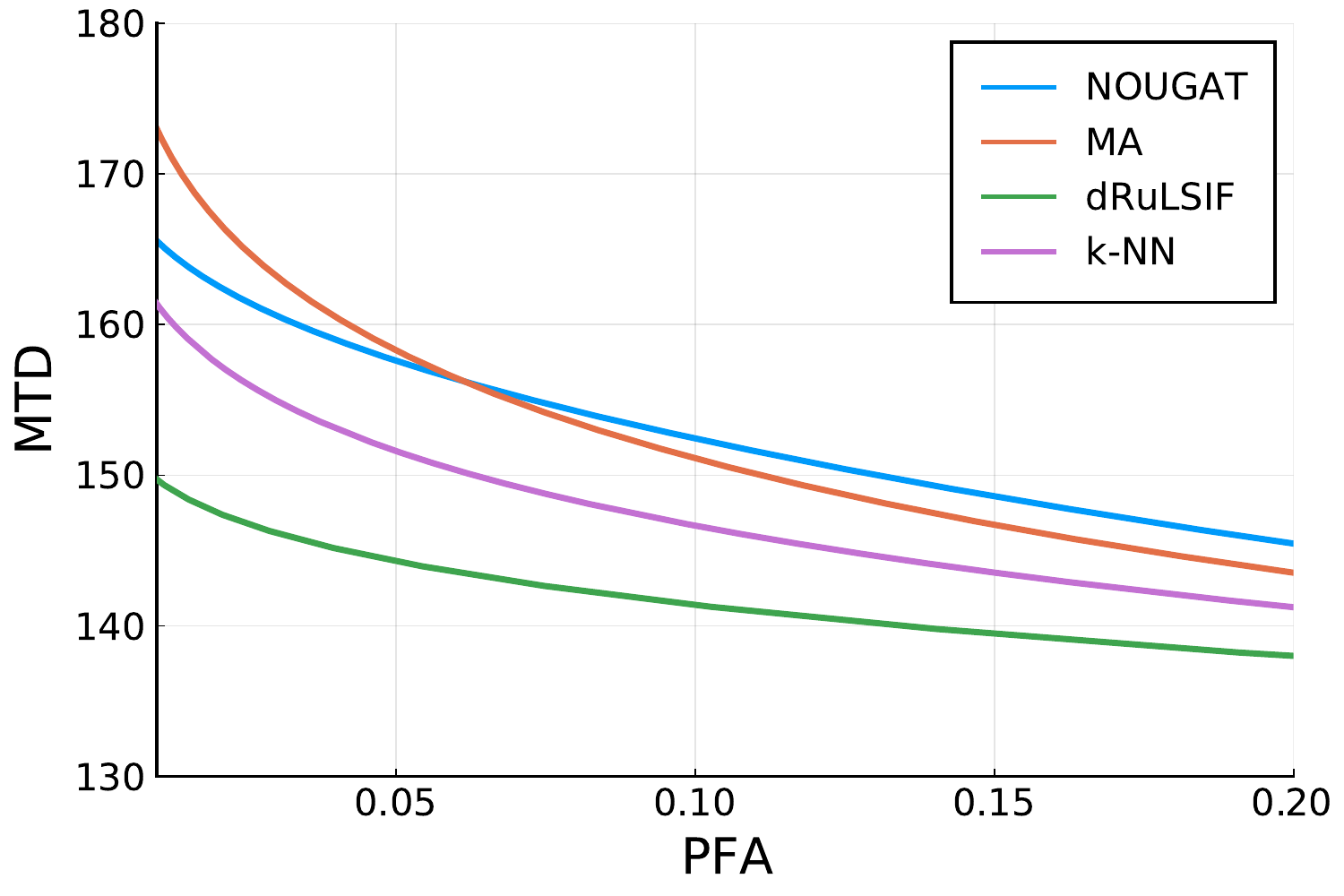}
	\caption{MTD as a function of PFA for NOUGAT, MA and dRuLSIF.}\label{MTD}
	\end{center}
\end{figure}

\begin{figure}
	\begin{center}
	\includegraphics[width=.8\columnwidth]{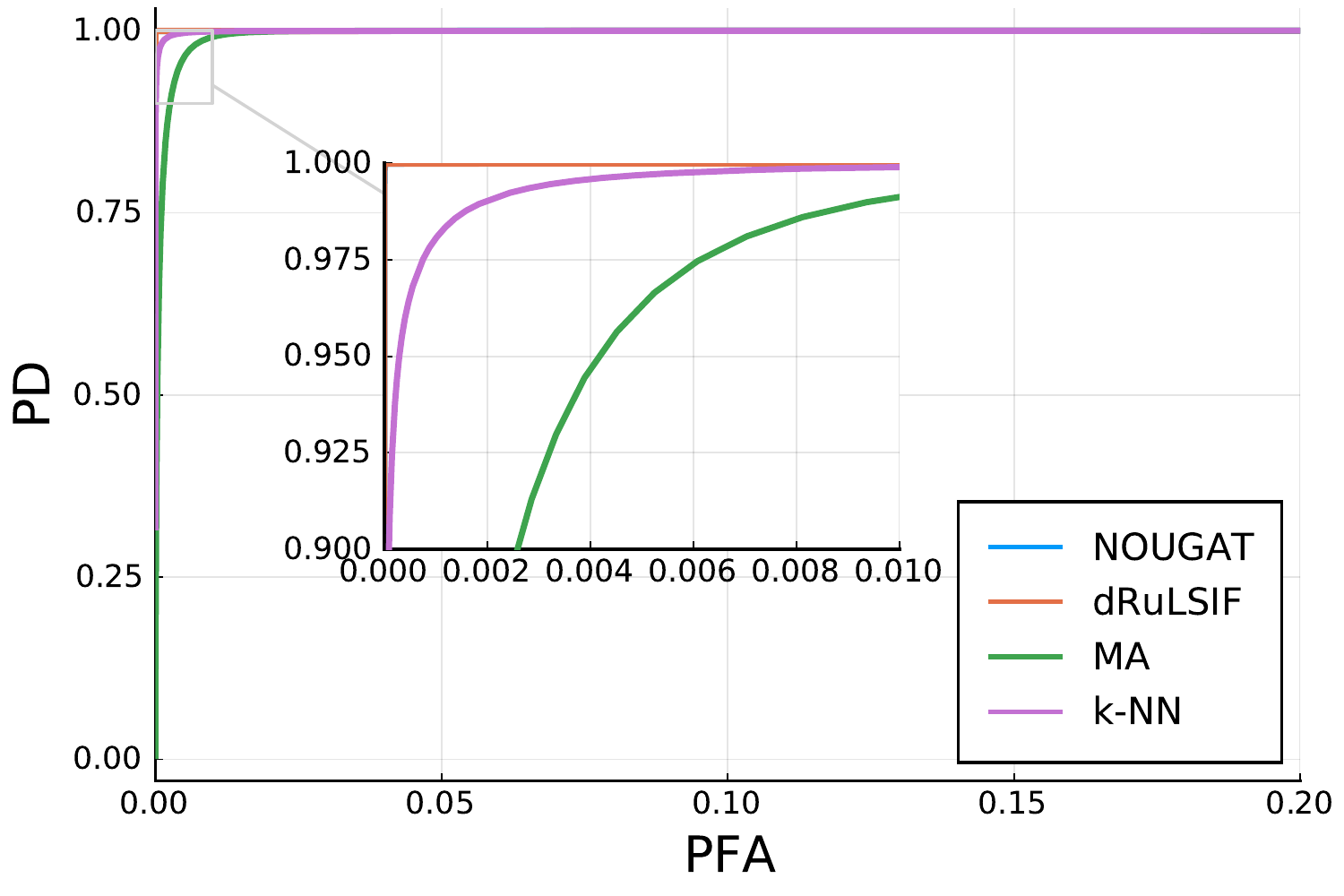}
	\caption{ROC curve for NOUGAT, MA and dRuLSIF.}\label{roc}
	\end{center}
\end{figure}

Figures \ref{MTFA} and \ref{MTD} provide the MTFA and MTD as a function of the Probability of False Alarm (PFA). The PFA was computed, for each algorithm, as the probability to detect an event at a time instant $t_a$ with $t_a<t_0$. The Probability of Detection (PD) was estimated as the probability to detect \emph{at least} a change at a time instant $t_a$ with $t_0 \leq t_a \leq n_t$, i.e. the probability that the test statistics is larger than the threshold at least once. Figure~\ref{roc} provides the Receiver Operating Characteristic (ROC) for the three algorithms.

Figure~\ref{MTFA} shows that, for $\text{PFA}>0.2$, the MTFA for the four algorithms is smaller than 40 samples. This means that the detection thresholds are too small and make the algorithms non-operational due to numerous false alarms. 

Focusing on the case $\text{PFA}<0.2$, we observe in Figure~\ref{roc} that when $\text{PFA}<0.01$, as expected from Figure~\ref{Statistics}, MA and k-NN reach the worst performance: for a given PFA, their PD are the smallest. We note that for $0.01<\text{PFA}<0.2$ the PDs of NOUGAT, dRuLSIF and k-NN are almost equal to 1. 

Figure~\ref{MTFA} shows that the MTFAs are almost the same with a larger delay of 5 samples for k-NN. Figure~\ref{MTD} shows that NOUGAT MTD is approximately 10 samples larger than dRuLSIF and 5 samples larger than k-NN. The delay of NOUGAT, which can be observed on Figure~\ref{Statistics} is, as explained before, due to the online update of NOUGAT. It is worthy to note that the performance of NOUGAT depends on $\mu$ and a smaller value would result in a smaller MTD.

\subsubsection{Computational cost} \label{Computational cost}

This experiment aims at comparing the computational cost of the four algorithms. It is worthy to note that, as long as the averages on the reference and test windows required by dRuLSIF, NOUGAT and MA are computed recursively, their computational cost does not depends on $N_\text{ref}$ and $N_\text{test}$. The data dimension $k$ only intervenes when computing $\bkappa_\omega(\cdot)$ and penalizes equally the three algorithms. On the contrary, the computational cost of k-NN strongly depends on the size of the windows, typically $O((N_\text{ref}+N_\text{test})\log(N_\text{ref}+N_\text{test})$ using a KD tree \cite{Bentley_1975}.

\begin{figure}
	\begin{center}
	\includegraphics[width=.8\columnwidth]{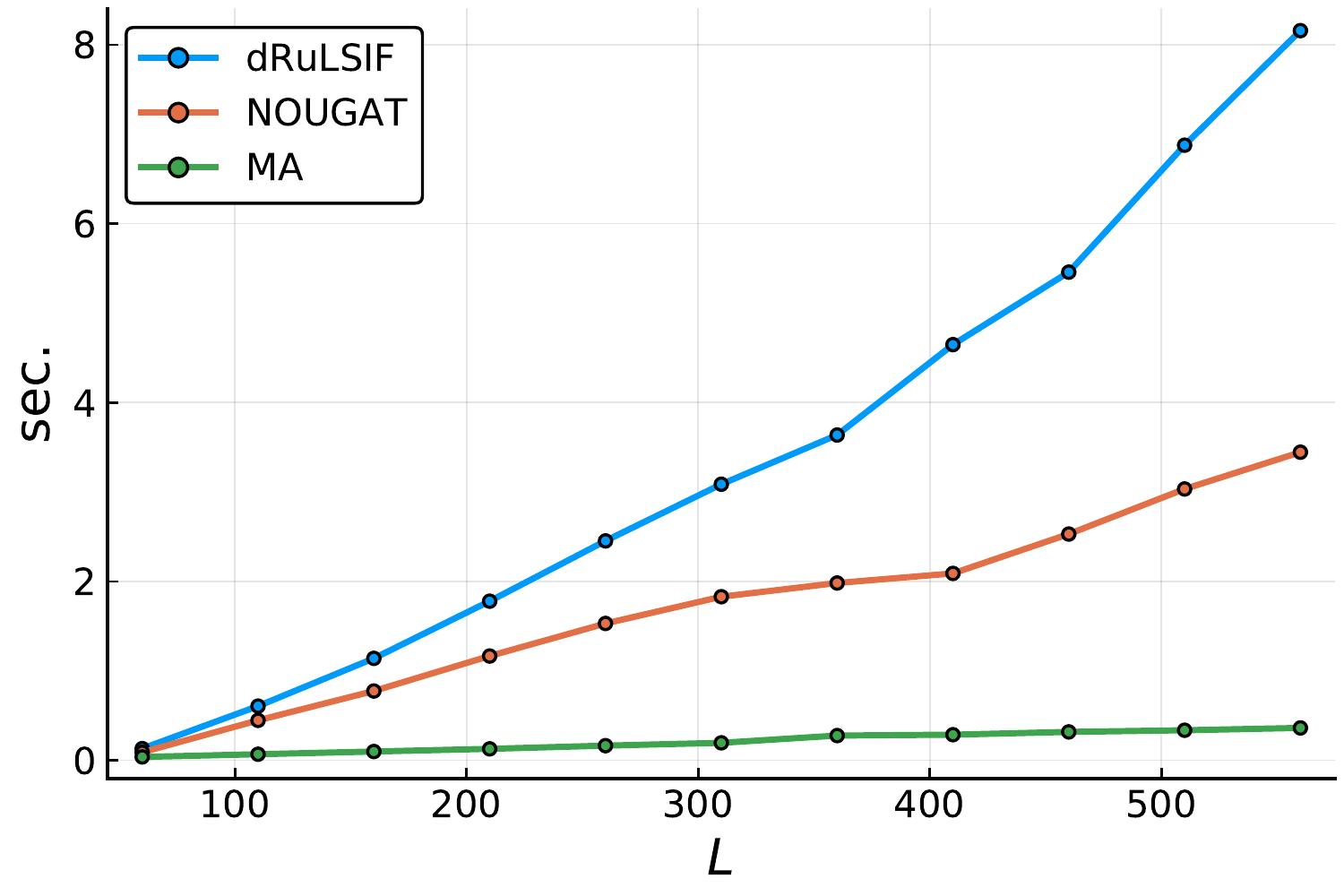}
	\caption{Run time for $n_t =1200$ samples as a function of the dictionary size $L$.}\label{timing}
	\end{center}
\end{figure}

Figures~\ref{timing} depicts the run time for the three kernel-based algorithms as a function of the dictionary size $L$, when processing $1200$ samples of dimension $k=6$. For each value of $L$, the run time was calculated as the median value of 1000 runs on an Intel Core i7 3,5 GHz. Figures~\ref{timing} shows that, compared to dRuLSIF, NOUGAT enjoys a considerably smaller run time while ensuring a good level of performance. 

\subsection{Experiments with real data}\label{Experiments}
\subsubsection{Credit card fraud detection}

The data set used in this experiment, called ``Credit Card Fraud Detection'', contains the 28 principal components of transactions made by European cardholders in September 2013. The data set is highly unbalanced as it contains 492 frauds out of 284,807 recorded transactions; see~\cite{Kaggle_Fraud} for more details. We chose to keep only 2,000 genuine transactions, and we inserted the 492 frauds in order to create two change-points at $t_0=1000$ and $t_0=1492$ in data stream $\{\by_t\}_{t\in \mathbb{N}}$. The four most significant principal components were used as inputs ($k=4$). The Gaussian kernel with kernel bandwidth $\sigma^2=14$, and reference and test windows of length $N_\text{ref}=N_\text{test}=114$, were considered for all algorithms. A regularization term with $\nu = 10^{-2}$ was used for NOUGAT and dRuLSIF, and the step size of NOUGAT was set to $\mu = 0.28$.

The online dictionary update procedure described in Section~\ref{sec:online} was used for all algorithms. The coherence threshold was set to $\eta_0 = 0.7$, leading to a dictionary size of $L = 100$.  Parameter vector $\btheta_{-1}$ was initially set to zero for NOUGAT.

\begin{figure}[!t]
	\begin{center}			\includegraphics[width=\columnwidth]{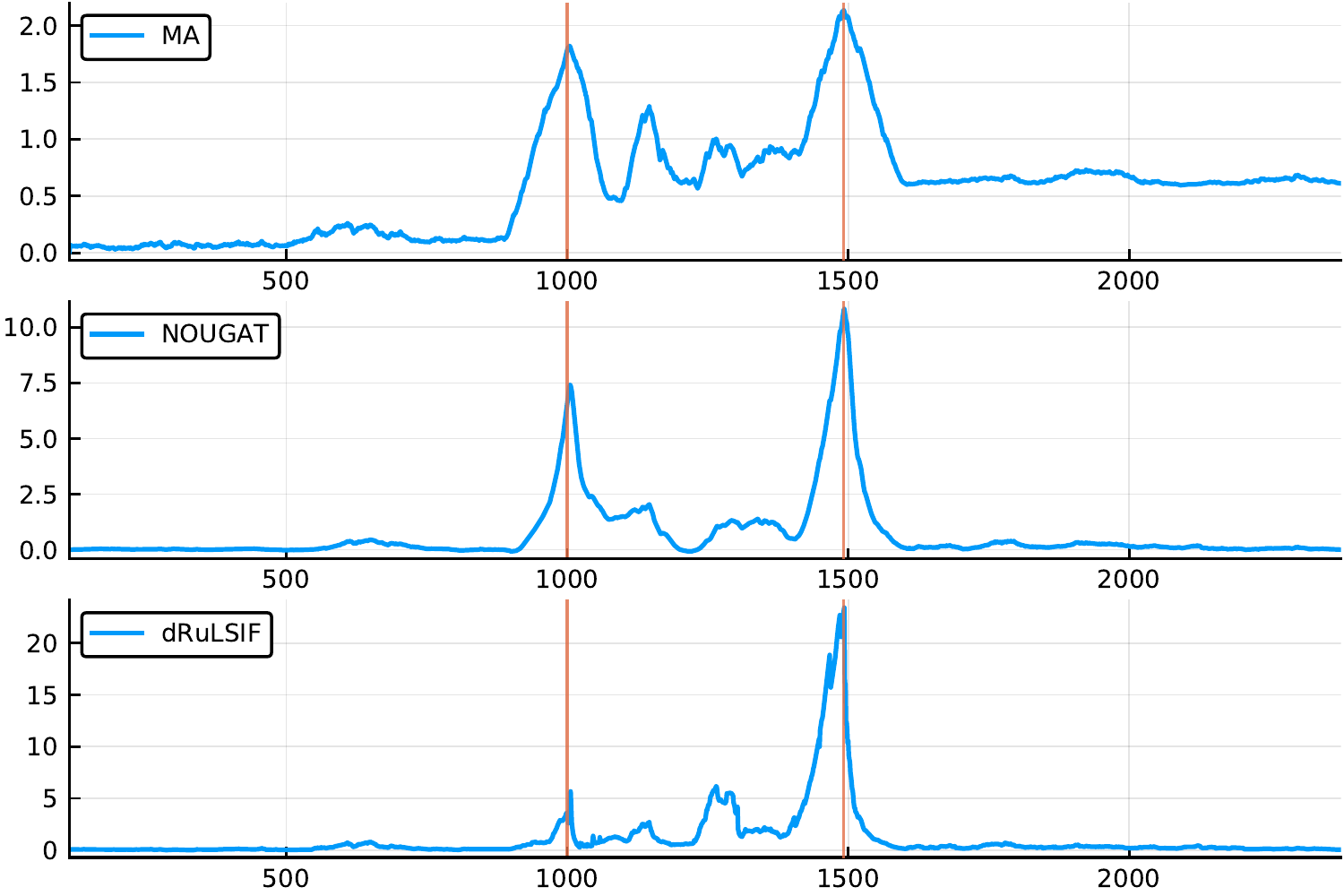}
	\end{center}
	\caption{Credit card fraud detection. The red lines correspond to $t_0+N_\text{test}$.}\label{Fig_fraud_detection}
\end{figure}

The results provided in Figure~\ref{Fig_fraud_detection} show that all the algorithms were able to detect the change-points. As expected, the detected change-points defined by the maximum value of each peak of the test statistics, were all in the vicinity of $t_0+N_\text{test}$.  Nevertheless, if MA was able to detect the two change-points marked by red lines in addition to some false positive detections, it suffered from a bias that deviated its static from zero after the first change-point. dRuLSIF hardly detected the first change-point, but successfully detected the second one. NOUGAT detected both change-points with less fluctuations of its detection statistics. Finally, NOUGAT and dRuLSIF test statistics fluctuated around 0 under the null hypothesis. These results  highlight the ability of the proposed algorithm to detect consecutive change-points.

\subsubsection{Sentiment change detection in Twitter data streams}

The data set used in this paragraph, called ``Twitter US Airline Sentiment'', is available at~\cite{Kaggle_tweet}. This data set contains tweets related to US Airline in February 2015, manually tagged as positive, negative and neutral. Raw tweets were first cleaned from non-ASCII characters. Stop words from Natural Language Toolkit (NLTK) corpus were also removed. Finally, tweets were represented, using a frequency-based method, in a linear space of dimension $k=50$. The series $\{\by_t\}_{t\in \mathbb{N}}$ was obtained by concatenating the 9178 negative-tagged tweets, the 2363 positive-tagged tweets and the 3099 neutral-tagged tweets.  Parameters were set to: $\mu = 10^{-1}$, $\nu = 5.10^{-33}$, and $N_\text{ref}=N_\text{test}=100$. A Gaussian kernel with $\sigma^2 = 1.3$ was used, along with an online dictionary learning procedure with a maximal coherence of $\eta_0 =10^{-3}$. This resulted in a dictionary of size $L=12$. Parameter vector $\btheta_{-1}$ was set to zero for NOUGAT.   
\begin{figure}
	\begin{center}	
	\includegraphics[width=\columnwidth]{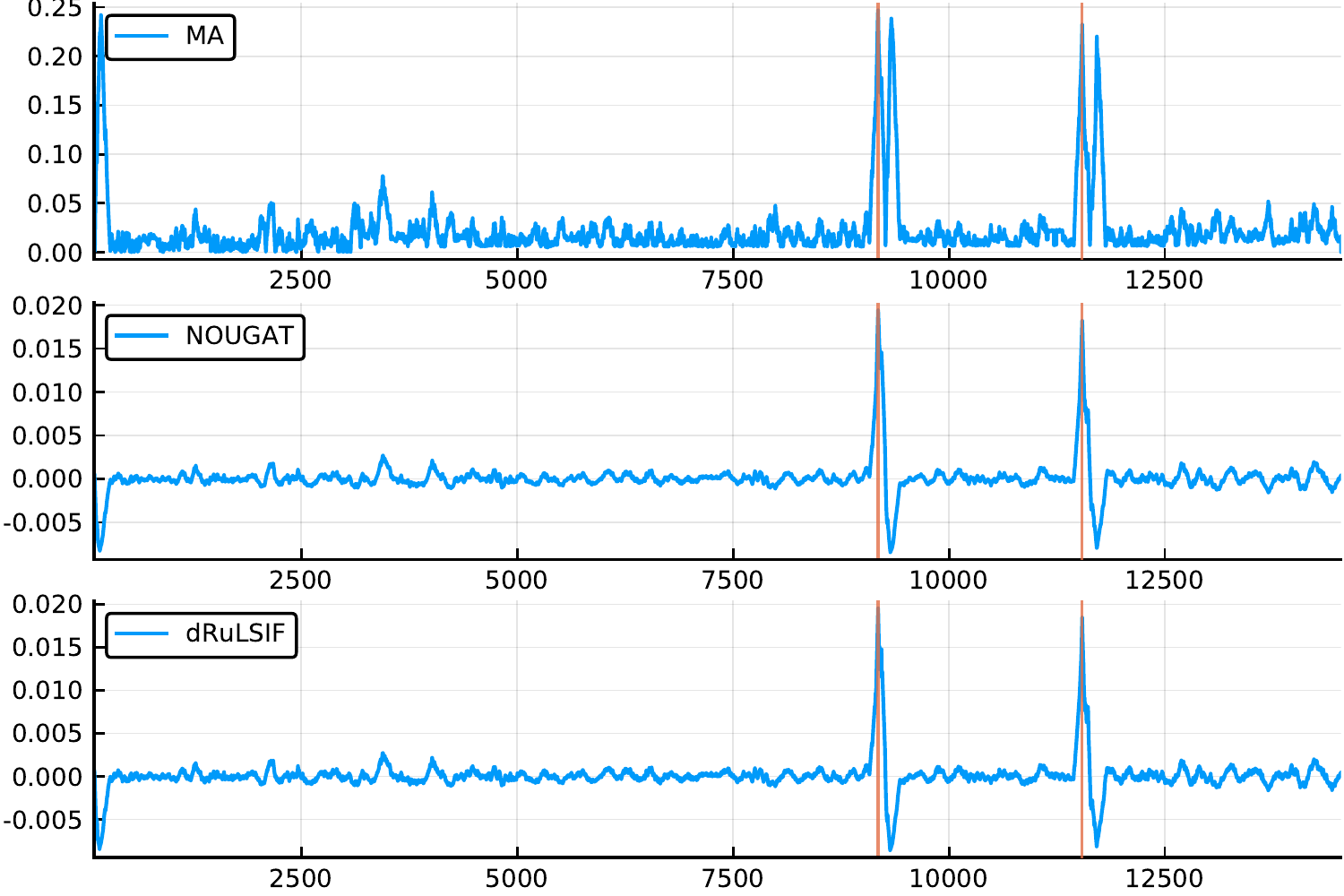}
	\end{center}
	\caption{Sentiment change detection in Twitter data stream. The red lines correspond to $t_0+N_\text{Test}$. \label{Sentiment_change_detection}}
\end{figure}

Figure~\ref{Sentiment_change_detection} provides the detection statistics of MA, NOUGAT and  dRuLSIF. MA produced 2 false alarms and the variance of its statistics was larger than the other two methods. NOUGAT and dRuLSIF led to similar results. Note that, as expected, for the three methods, the peak at the first (negative/positive) transition was slightly higher than the peak at the second (positive/neutral) transition.

\subsubsection{Change detection in satellite telemetry}

The data set used in this experiment was provided by Thales Alenia Space. It consists of an electrical current signal produced by a panel of a geostationary satellite. The sampling period of data points is approximately 32 seconds, and the data span a time period of six months. 
A change point is known to occur at time instant $t_0= 177,630$. Marked by a red line, it represents a drop in the quantity of electrical current produced by the panel due to the loss of solar cells.

\begin{figure}
	\begin{center}			
	\includegraphics[width=\columnwidth]{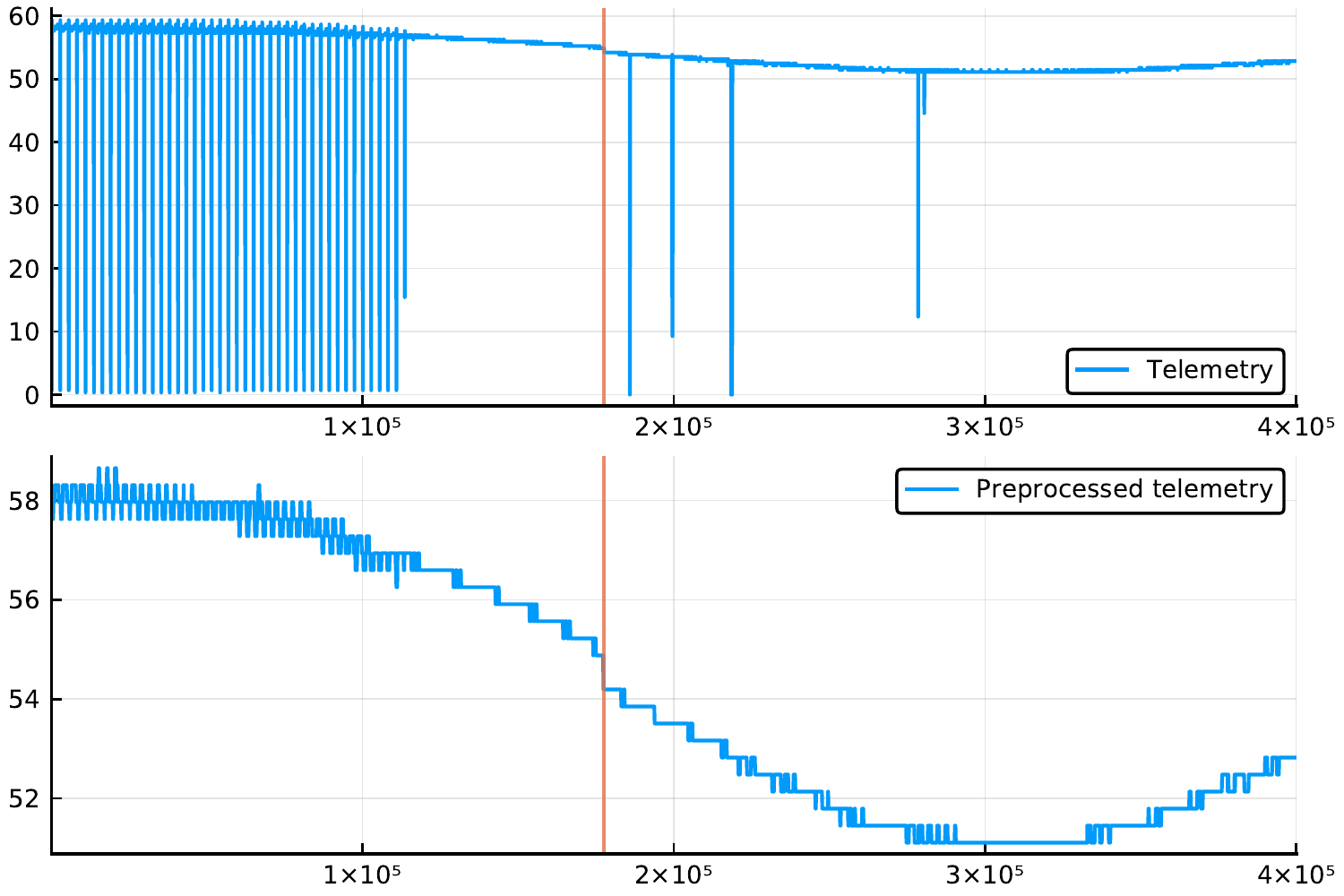}
	\caption{Top: Telemetry. Bottom: Signal pre-processed by a median filter.}\label{TAS1}
\end{center}
\end{figure}

Figure~\ref{TAS1}~(top) partly shows the electrical current signal. The consecutive current drops observed at the beginning of the signal represent each a period of eclipse. These drops were removed using a median filter of length $600$, which corresponds to the maximum duration of an eclipse. The filtered signal is shown in Figure~\ref{TAS1}~(bottom). 
Vectors $\by_t$ of dimension $k = 10$ used as inputs for the detection algorithms were extracted using a sliding window as explained in~\eqref{sigtovec}. Window lengths $N_\text{ref} =N_\text{test} = 3000$ were used. This value corresponds approximately to a 1-day period, which is sufficient to capture the main stationary characteristics of the signal. These characteristics depend on changes in the distance from the panels to the Sun, and the angle of incidence of the sunlight.

\begin{figure}
	\begin{center}
	\includegraphics[width=\columnwidth]{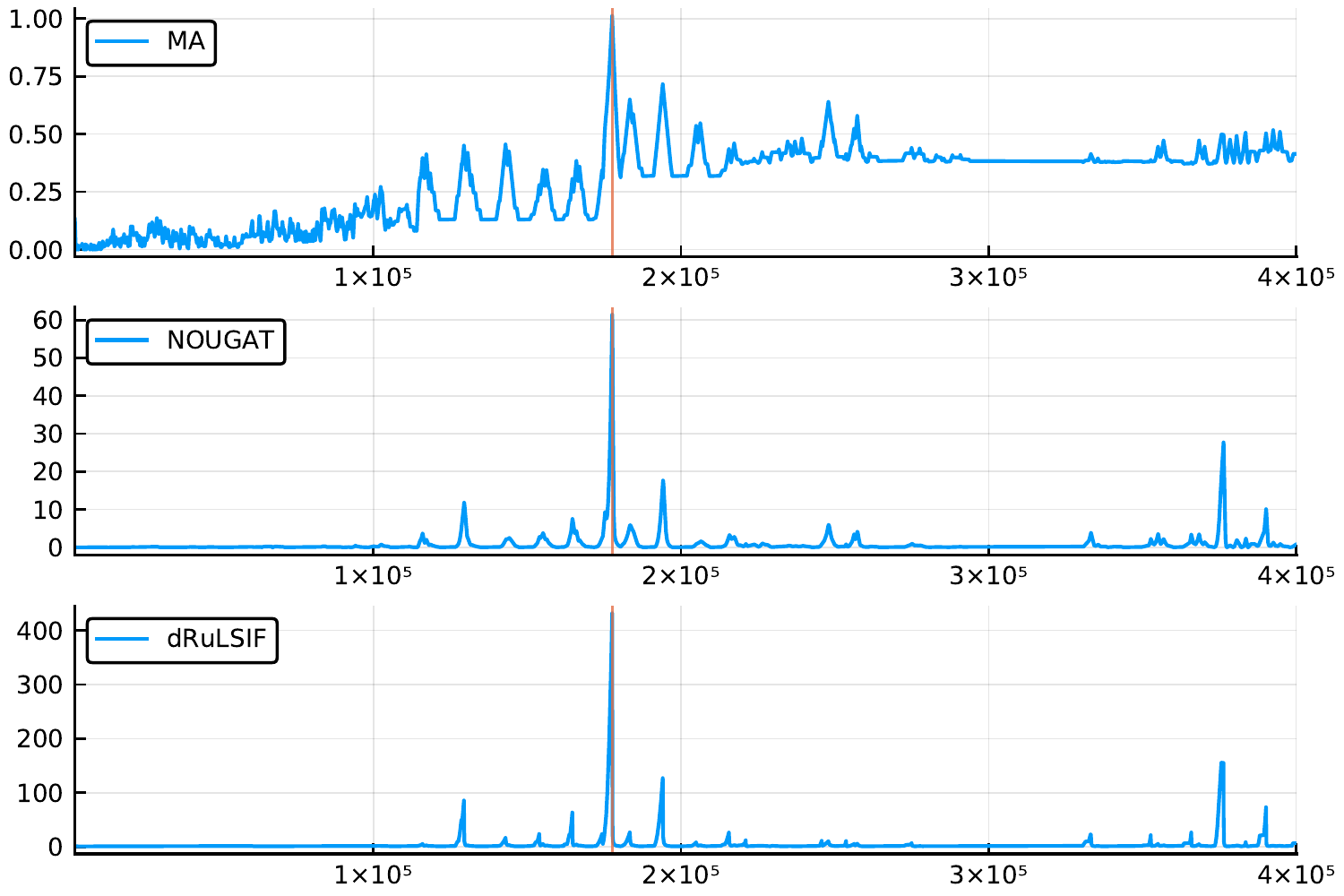}
\caption{CPD in satellite telemetry data.
The red line corresponds to $t_0+N_\text{Test}$.}\label{TAS3}
	\end{center}
\end{figure}
An online dictionary learning procedure was used with a maximal coherence value of $0.5$. This resulted in a dictionary of size $L=33$. The three algorithms produced false alarms. MA had a bias, dRuLSIF and NOUGAT showed similar results but with a much lower computational load for NOUGAT. Note that computational load is a key concern for this application.

\section{Conclusion} \label{Conclusion}

We introduced an online kernel-based change-point detection method built upon direct estimation of the density ratio on consecutive time intervals. We analyzed its behavior in the mean and mean square sense. Finally, we evaluated its detection performance and we compared it to state-of-the-art kernel-based methodologies, MA and RuLSIF, and to another approach based on k-NN. We showed that our algorithm has a considerably lower computational complexity than dRuLSIF while ensuring comparable performance. Experiments on real-world data proved the usefulness and efficiency of our algorithm in a number of applications. These applications involved different types of data, namely, text data, raw data, and features extracted from data, showing the interest in using non-parametric techniques to perform change-point detection

We leave for future work the derivation of methods for kernel selection, and the opportunity of using a symmetric detection statistic where covariance information on the test interval would also be considered.

\appendix

\section{Computation of $\bH$ and $\bh$}
\label{Computation of H and h}

Considering the Gaussian reproducing kernel:
\begin{equation*}
	\label{gauss_kern}
	\kappa (\by , \by') = e^{-\frac{\| \by -\by' \|^{2}}{2\sigma ^{2}}}
\end{equation*}
the entries of \eqref{esp_H} are given by:
\begin{align*}
&[\bH]_{\ell,q}  = e^{-\frac{\|\by_{\omega_\ell}\|^2+\|\by_{\omega_q}\|^2}{2\sigma^2}}\;\; \esp_{p_\text{ref}(\by)}\left\{e^{-\frac{\|\by\|^2- (\by_{\omega_\ell} +\by_{\omega_q})^\top\by}{\sigma^2}}\right\}
\end{align*}
and those of $\bh$ by:
\begin{align*}
[\bh]_\ell &= e^{-\frac{\|\by_{\omega_\ell}\|^2}{2\sigma^2}}\;\esp\left\{e^{-\frac{\|\by\|^2- 2\by_{\omega_\ell}^\top \by}{2\sigma^2}}\right\} \end{align*}
with $\ell$, $q\in\{1,\ldots, L\}$. These expectations can be computed for Gaussian distributed entries $\by_i \sim \mathcal{N}(\bmu, \bR) $ using the moment generating function of a quadratic form of a Gaussian vector~\cite{omura65}:
\begin{align*}
&[\bH]_{\ell,q}  = e^{-\frac{\|\by_{\omega_\ell}\|^2+\|\by_{\omega_q}\|^2}{2\sigma^2}}~\Psi \big(\frac{-1}{\sigma^2},\bI,-(\by_{\omega_\ell}+\by_{\omega_q}),\bmu,\bR \big)\\
&[\bh]_\ell = e^{-\frac{\|\by_{\omega_\ell}\|^2}{2\sigma^2}} ~\Psi\big(\frac{-1}{2\sigma^2},\bI,-2 \by_{\omega_\ell},\bmu,\bR\big)
\end{align*}
where:
\begin{align}
&\Psi(s, \bW, \bb, \bmu, \bR) \\&= |\bI- 2s \bW \bR|^{-\frac{1}{2}}\exp\Big(s\left[(\bmu^\top \bW \bmu + \bb^\top \bmu)\right. \nonumber\\                 &+ \left. \frac{s}{2}\| 2\bW \bmu + \bb\|^2_{\bR(\bI - 2s\bW \bR)^{-1}}\right] \Big) \label{Psi}
\end{align}

\section{Computation of $\bGamma$ and $\bDelta$}\label{Computation of Gamma and Delta}

The  $(r,s)$-th entry of the $(q,n)$-th block of matrix $\bGamma$, that is, $\esp \{\kappa_{\omega_q}(\by_{i})\kappa_{\omega_n}(\by_{i})  \bkappa_{\bomega}(\by_{i})\bkappa_{\bomega}(\by_{i})^\top \}$ is given by: 
\begin{align}
&\bGamma_{(q-1)L +r, (n-1)L +s} =
e^{-\frac{\|\by{\omega_q}\|^2+ \|\by_{\omega_n}\|^2+\|\by_{\omega_r}\|^2+\|\by_{\omega_s}\|^2}{2\sigma^2}}\nonumber\\
&\Psi \Big(\frac{-1}{\sigma^2}, ~2\bI,~-(\by_{\omega_q}+\by_{\omega_n}+\by_{\omega_r}+\by_{\omega_s}),~\bmu,~\bR \Big)    
\end{align}

Similarly, the $r$-th entry of the $(q, n)$-th block of $\bDelta$, that is, $\esp \{\kappa_{\omega_q}(\by_{i})    \kappa_{\omega_n}(\by_{i})\bkappa_{\bomega}(\by_{i})\}$, is given by:
\begin{align}
&\bDelta_{(q-1)L +r, n} = e^{-\frac{\|\by_{\omega_q}\|^2 + \|\by_{\omega_n}\|^2+\|\by_{\omega_r}\|^2}{2\sigma^2}}\nonumber\\
&\Psi \Big( \frac{-1}{2\sigma^2},~3\bI,~-2(\by_{\omega_q}+\by_{\omega_n}+\by_{\omega_r}),~\bmu,~\bR \Big)    
\end{align}

\balance

\section*{Acknowledgement}
This work has been supported by the French government, through the 3IA Côte d’Azur Investments in the Future project managed by the National Research Agency (ANR) with the reference number ANR-19-P3IA-0002.

Ikram Bouchikhi was partly funded by Thales Alenia Space TAS.

\bibliographystyle{elsarticle-num} 
\bibliography{biblio}

\end{document}